\documentclass[aps,pra,reprint,amssymb,longbibliography,preprintnumbers,twocolumn,amsmath,superscriptaddress,floatfix]{revtex4-2}
\usepackage{graphicx}
\usepackage{bm}
\usepackage{hyperref}
\usepackage{physics}
\usepackage{orcidlink}
\usepackage[final]{showlabels}

\usepackage{color}
\usepackage[normalem]{ulem}

\newcommand{\be}{\begin{equation}}
	\newcommand{\ee}{\end{equation}}
\newcommand{\ba}{\begin{eqnarray}}
	\newcommand{\ea}{\end{eqnarray}}
\newcommand{\bs}{\begin{subequations}}
	\newcommand{\es}{\end{subequations}}
\newcommand{\bw}{\begin{widetext}}
	\newcommand{\ew}{\end{widetext}}

\renewcommand{\vec}[1]{\boldsymbol{\mathbf{#1}}}
\newcommand{\tens}[1]{\overleftrightarrow{#1}}

\begin{document}

\title{Theoretical description of high-harmonic spectroscopy for uniaxial nematic and isotropic liquid crystals}

\author{Simon Vendelbo Bylling Jensen \orcidlink{0000-0002-6749-0978}}
\affiliation{Max Planck Institute for the Structure and Dynamics of Matter and Center for Free-Electron Laser Science, Hamburg 22761, Germany}

\author{Marta L. Murillo-Sánchez \orcidlink{0000-0002-9681-349X}}
\affiliation{Max-Planck-Institut für Kernphysik, Saupfercheckweg 1, D-69117 Heidelberg, Germany}

\author{Andrea Annunziata~\orcidlink{0000-0002-7599-6338}}
\affiliation{Max-Planck-Institut für Kernphysik, Saupfercheckweg 1, D-69117 Heidelberg, Germany}

\author{Angel Rubio \orcidlink{0000-0003-2060-3151}}
\affiliation{Max Planck Institute for the Structure and Dynamics of Matter and Center for Free-Electron Laser Science, Hamburg 22761, Germany}
\affiliation{Initiative for Computational Catalysis (ICC), The Flatiron Institute, New York, New York 10010, USA}

\author{Laura Cattaneo \orcidlink{0000-0001-7492-3850}}
\affiliation{Max-Planck-Institut für Kernphysik, Saupfercheckweg 1, D-69117 Heidelberg, Germany}

\author{Nicolas Tancogne-Dejean \orcidlink{0000-0003-1383-4824}}
\affiliation{Max Planck Institute for the Structure and Dynamics of Matter and Center for Free-Electron Laser Science, Hamburg 22761, Germany}
\affiliation{Research Center Future Energy Materials and Systems of the University Alliance Ruhr and Interdisciplinary Centre for Advanced Materials Simulation, Faculty of Physics and Astronomy, Ruhr University Bochum, Universitätsstraße 150, D-44801 Bochum, Germany }

\date{\today}

\begin{abstract}
Enabled by their controllable optical anisotropy, liquid crystals form a cornerstone of linear optical and display technologies. Bringing the versatile platform into the nonlinear regime, we explore its nonlinear optical anisotropy by high-order harmonic generation spectroscopy. Through a comprehensive study targeting gas, liquid, nematic, and crystalline phases, we present a semi-empirical tight-binding approach that provides insight into the intermolecular character of ultrafast light-driven electron dynamics across soft-matter mesophases. We demonstrate how intermolecular electron hybridization enables spectroscopic access to structural properties, such as the nematic director and order parameter.
Outlining such foundational mechanisms opens avenues towards nonlinear liquid crystal devices and ultrafast soft-matter spectroscopies. 
\end{abstract}  
\maketitle

\section{Introduction}
Liquid crystals (LCs) are soft-matter systems that combine fluidity with a degree of molecular order, bridging the gap between liquids and solids. 
In most of the cases, they consist of uniaxial rod- or disk-like molecules that can translate, tumble, or rotate while attaining certain degrees of macroscopic structural order and yielding specific crystalline features such as optical birefringence or other electrical, magnetic, and optical anisotropic properties~\cite{Goodby2016}. They exist in various thermodynamic mesophases with different structural orders, many of which consist of partially aligned molecules along a common axis called the director, such as what governs the nematic phase. Some are further arranged into layers to form \textit{e.g.} the smectic phase, achieving even higher degrees of combined positional and orientational order.
Crucially, thermotropic LCs exhibit controllable structural order, as their macroscopic structural mesophase can be tuned by temperature~\cite{Goodby2016}. Further, for partially ordered phases such as the nematic phase, the microscopic molecular structure, \textit{i.e.} the orientation of the molecular director, can be controlled by electric fields. Their tunable structural and optical properties have been employed with decades of expertise in linear optics applications such as the multibillion-dollar display technology industry~\cite{ANDRIENKO2018520}. 
Advancing the LC technologies from the linear towards the nonlinear optical regime gives prospects for exploiting controllable material structure to open a playground for rich light-driven electron dynamics to achieve innovative controllable nonlinear optical devices. 

During the last decades, strong-field light-driven electron dynamics have given rise to non-perturbative high-order harmonic generation (HHG) across a wide range of systems originating from atomic and molecular gases~\cite{McPherson:87,M_Ferray_1988} towards metals~\cite{10.1063/1.89628,PhysRevLett.46.29,Korobenko2021,PhysRevLett.76.1832}, semiconductors~\cite{Ghimire2011}, insulators~\cite{PhysRevA.52.R25,Luu2015}, nanostructures~\cite{Han2016}, monolayers~\cite{doi:10.1126/science.aam8861}, Dirac semimetals~\cite{PhysRevLett.124.117402}, doped semiconductors~\cite{10.1063/5.0047421}, Weyl semimetals~\cite{Lv2021}, topological insulators~\cite{Bai2021}, rare noble-gas solids~\cite{Ndabashimiye2016},  amorphous solids~\cite{You2017}, liquids~\cite{Luu2018}, and molecular crystals~\cite{Wiechmann2025}.
Across such systems, the microscopic generation mechanism is often described in terms of a three-step model, consisting of ionization/excitation, propagation, and recollision/recombination, in which electrons can acquire energy from a strong driving field to emit a nonlinear higher harmonic response~\cite{osti_10169672,PhysRevLett.70.1599,PhysRevA.49.2117,PhysRevLett.113.073901}. 

Most recently, the HHG process in liquid systems has gained increased interest due to the observation of a relatively constant maximal cut-off energy in the spectral response across different wavelengths, intensities, and compounds. A striking contrast to the well-established quadratic and linear electric field strength scaling of, respectively, gas and solid state targets~\cite{PhysRevLett.68.3535,Ghimire2011,Luu2018,PhysRevResearch.5.043030,Mondal2023}. 
A microscopic interpretation was proposed, and corroborated by experiments, considering a collection of single molecular emitters, from which the most energetic electron trajectories could be attenuated due to scattering off neighboring molecules~\cite{Mondal2023}.

When advancing to the more complex systems of liquid solutions, interpretations based on single-molecular emitters again showed success, as some solutions are well described as a simple sum of the solute and solvents harmonic yields~\cite{doi:10.1073/pnas.2514825122}. For solutions with characteristic localized solvation shells, a strong suppression of specific harmonics was interpreted as a sign of electron scattering off molecules in the solvation shell.
In contrast, a pronounced second spectral plateau was also observed and interpreted as arising from electrons that recombine at neighboring molecules~\cite{Mondal2026}. Such interpretation requires the electronic states in question, the highest-occupied molecular orbital (HOMO) and lowest unoccupied molecular orbital (LUMO), to be delocalized across full solvation shells, reminiscent of the delocalized Bloch state interpretation of condensed matter systems. 
Intermolecular interactions thereby attain a contradictory role, both attributed to suppressing harmonic generation through scattering and dephasing, and also offering additional channels for electron recombination. To elucidate the fundamental role of intermolecular interactions, LCs provide an ideal platform, due to the unique possibility to shape and control the microscopic structural properties on demand. Furthermore, the anisotropic optical properties of LCs hold great potential to provide additional spectroscopic information about the fundamental microscopic electron processes at hand. Optical anisotropy has, for instance, historically enabled tomographic imaging of molecular orbitals~\cite{Itatani2004,Vozzi2011}, tracking attosecond nuclear motion~\cite{doi:10.1126/science.1123904} and resolving ultrafast electronic charge migration~\cite{doi:10.1126/science.aab2160} in optically aligned gaseous targets. Additionally, for solids, optical anisotropies have been utilized to probe crystal structure symmetries~\cite{Ghimire2011,Schubert2014}, unravel anisotropic couplings and curvatures of the energy band structure~\cite{PhysRevLett.120.243903,PhysRevA.96.063412} and distinguish between underlying driven-electron or exciton mechanisms~\cite{Langer2016}. LCs represent a natural extension of this paradigm, offering a highly tunable anisotropic environment where nonlinear optical processes can be tailored through macroscopic order. 

Historically, the nonlinear optics landscape of nematic LCs has been characterized by two distinct mechanisms: the \textit{nonelectronic} collective reorientation of the molecular director and the \textit{electronic} polarizability of individual molecules~\cite{KHOO2009221,doi:https://doi.org/10.1002/9783527671403.hlc139,khoo2014multiple}. While the former provides \textit{giant} optical nonlinearities, it is inherently limited by relatively slow response times in the millisecond to nanosecond range~\cite{10.1063/1.457034}.
In contrast, electronic nonlinearities enable ultrafast responses on picosecond and femtosecond timescales, allowing for the direct probing of molecular structure and symmetry~\cite{PhysRevA.30.2118,KHOO2009221}. Perturbative nonlinear experiments have extensively utilized these electronic responses to characterize different LC phases. Second harmonic generation, for instance, has served as a sensitive probe for symmetry-breaking in ferroelectric LCs~\cite{Ferroelectric} and for resolving the structure of freely suspended films where centrosymmetry is lifted~\cite{PhysRevA.34.4303}.
Similarly, third harmonic generation has been employed to investigate electronic hyperpolarizabilities and molecular alignment~\cite{PhysRevA.34.5051,10.1063/1.1763220}, as well as the enhancement of nonlinear effects in nonlocal spatial solitons~\cite{Peccianti:10}. These processes are deeply influenced by the specific molecular symmetry; for example, the low C$_2$ symmetry characteristic of the nCB series allows for a complex interplay between molecular orientation and orbital-driven nonlinearities. Furthermore, intense laser fields can via the molecular polarizability induce sub-nanosecond order dynamics, involving both uniaxial and biaxial modes of the dielectric tensor~\cite{PhysRevLett.111.107802}. These studies highlight how the perturbative response is intrinsically tied to the interplay between the macroscopic director and microscopic electronic transitions.

\begin{figure} \includegraphics[width=\columnwidth]{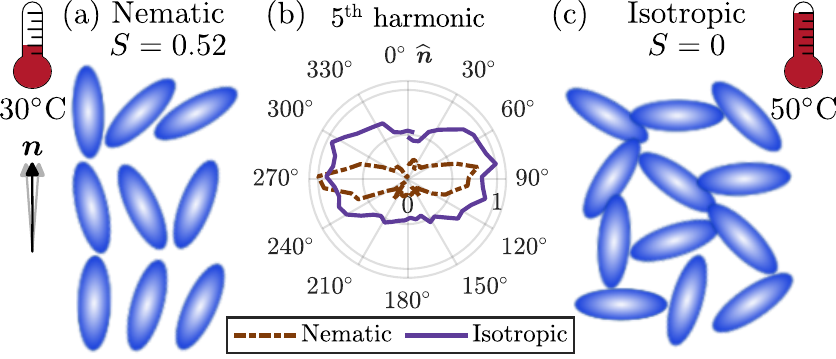}
	\caption{Illustration of high-order harmonic spectroscopy of (a) nematic and (c) isotropic LC phases.  (b) The polarization dependency of the harmonic yield with respect to the nematic director ${\bm n}$ for the $5^\text{th}$ harmonic of 8CB, extracted from Ref.~\cite{10.1063/5.0191184}. Analogous characteristic polarization dependencies are found across various samples and harmonic orders, see Ref.~\cite{10.1063/5.0191184} for details.} \label{fig:1}
\end{figure}

In this paper, we unravel the fundamental role of microscopic intermolecular electron dynamics that govern the nonlinear anisotropic optical response of LCs. We provide an interpretation of recent pioneering experiments of Ref.~\cite{10.1063/5.0191184}, see characteristic polarization dependent harmonic response in Fig.~\ref{fig:1}. Here, across several liquid crystal sample materials, the high-temperature liquid phase showcased a rather isotropic nonlinear response, while the lower temperature nematic phase displayed a preferential direction perpendicular to the molecular director. The observations for the nematic phase are contrary to the expected behavior for elongated independent molecular emitters within perturbative soft-matter optics, hinting at the appearance of a more evolved generation mechanism in the nonlinear regime. Through a comprehensive study targeting anisotropic spectroscopies across gas, liquid, nematic, and crystalline phases, we propose a first-principles-informed semi-empirical theoretical model that can help guide novel ultrafast all-optical spectroscopes of soft-matter systems. Together with the accompanying experimental study of Ref.~\cite{experimental_paper}, we seek to underpin the prospects of utilizing the LC platform to unravel the foundational light-driven electron dynamics in attosecond spectroscopies across various structurally ordered phases. With it, we seek to provide additional fundamental insight towards achieving spectroscopic sensitivity to the spatial hybridization of electronic wavefunctions in liquid or disordered environments, which could be extended to resolve relevant dynamical processes within chemistry or biology.

\section{Theoretical models and methods} \label{Section2}

\subsection{Rod-like approach}
We start by showcasing a simple perturbative optical approach to soft-matter systems in which we consider a collection of $N$ independent molecules in a macroscopic volume~\cite{RevModPhys.46.617}. This approach, while clearly not predictive, is based on symmetry arguments and aims at exploring how the statistical averaging over many independent molecules could give rise to general spectroscopic characteristics. For this, we consider a uniaxial LC made of non-polar elongated molecules. We define $\vec{\nu}^{(i)}$ the unit vector along the axis of the $i^\mathrm{th}$ molecule.
The ordering in a nematic LC is given by the rank-2 tensor
\begin{equation}
 S_{\alpha\beta} = \frac{1}{N}\sum_i (\nu_\alpha^{(i)}\nu_\beta^{(i)})-\frac{1}{3}\delta_{\alpha\beta} = \langle \nu_\alpha \nu_\beta \rangle -\frac{1}{3}\delta_{\alpha\beta} \,, \label{eq:1}
\end{equation}
where the average $\langle \nu_\alpha \nu_\beta \rangle$ is defined by the first part of the equation.
Here $\alpha$ and $\beta$ are Cartesian coordinates and $\delta_{\alpha\beta}$ is the Kronecker delta. The sum in the average is considered over all $N$ molecules in a macroscopic volume.
If we assume uniaxial molecular symmetry, the order parameter can be written in the form of
\begin{equation}
 S_{\alpha\beta} = S\left(n_\alpha n_\beta -\frac{1}{3}\delta_{\alpha\beta}\right)\,, 
 \label{eq:def2}
\end{equation}
where $n_\alpha$ are the components of the molecular director~$\vec{n}$~\cite{RevModPhys.46.617}.
Assigning $\vec{n}$ along $z$, it follows immediately from Eq.~\ref{eq:def2} that non-zero components of the tensor $S_{\alpha,\beta}$ are $S_{zz} = \frac{2}{3}S$, and $S_{xx}=S_{yy} = -\frac{1}{3}S$.
The scalar $S$ is a measure of the alignment of the molecules, denoted as the order parameter, and is usually defined for uniaxial nematic LCs as
\begin{equation}
 S = \int_0^\pi P_2(\cos(\theta)) f(\theta)\sin(\theta) d\theta = \langle  P_2(\cos(\theta)) \rangle \,,
\end{equation}
where $f(\theta)\sin(\theta) d\theta$ represents the fraction of molecules whose axes make angles between $\theta$ and $\theta+d\theta$, $P_2$ is the second Legendre polynomial, and the notation $\langle . \rangle$ indicate the averaging over the associated molecules.

\subsubsection{Linear response}
We first summarize how the optical properties of both a single molecule and the macroscopic LC naturally follows from the considerations above. More details can be found in reviews on LCs, see for instance Ref.~\cite{RevModPhys.46.617}.  For elongated, rod-like molecules, the polarizability tensor of the molecule possesses two non-zero components, which we refer to as $\alpha_{\parallel}$ and $\alpha_{\perp}$ for the response along and perpendicular to the molecular axis. These determines the polarizability tensor of the isolated molecule, that aligned along the $\hat{\vec{z}}$ direction, can be written as
\begin{equation}
 \tens{\alpha} = 
 \begin{pmatrix}
\alpha_{\perp} & 0 & 0\\
0 & \alpha_{\perp} & 0\\
0 & 0 & \alpha_{\parallel}
\end{pmatrix}\,, \nonumber
\end{equation}
where the frequency is omitted for conciseness. 
Given a more general direction of a single molecule $\vec{\nu}^{(i)}$, the components of the polarizability tensor can be written simply as
\begin{equation}
 \alpha_{ij}(\vec{\nu}) = \alpha_{\perp}\delta_{ij} + ( \alpha_{\parallel}-\alpha_{\perp})\nu_i\nu_j\,.
\end{equation}
Taking the average over an ensemble of molecules we find
\begin{equation}
 \langle\alpha_{ij}\rangle = \alpha_{\perp} \delta_{ij} + ( \alpha_{\parallel}-\alpha_{\perp}) \langle \nu_i\nu_j\rangle\,.
\end{equation}
Using Eq.~\eqref{eq:1} and Eq.~\eqref{eq:def2}, we obtain the averaged polarizability tensor in terms of the scalar order parameter~$S$
\begin{equation}
 \langle\alpha_{ij}\rangle = \big(\alpha_0-\frac{\alpha_a S}{3}\big) \delta_{ij} + S\alpha_a n_i n_j \,,
\end{equation}
where we have the average $\alpha_0 = \frac{2\alpha_{\perp} + \alpha_{\parallel}}{3}$ and the anisotropy component $\alpha_a = \alpha_{\parallel}-\alpha_{\perp}$ as Eq.~(15.3) in Ref.~\cite{RevModPhys.46.617}. We thus obtain that the average polarizability tensor corresponds to a uniaxial one, which is aligned along the director, with $\langle\alpha\rangle_{\perp} = \alpha_0-{\alpha_a S}/{3}$ and 
$\langle\alpha\rangle_{\parallel} = \alpha_0+{2\alpha_a S}/{3}$. 

\begin{figure} 
	\includegraphics[width=\columnwidth]{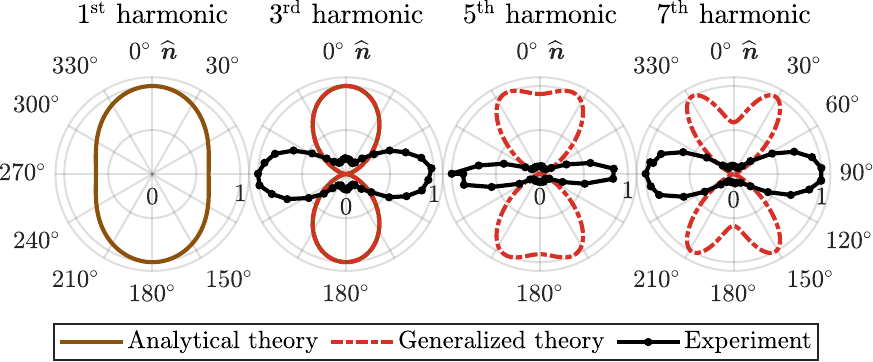}
	\caption{Polar plot of the polarization response of the nematic LC for $S=0.5$, at the fundamental laser frequency, as well as higher-order 3$^\text{rd}$, 5$^\text{th}$, and 7$^\text{th}$ harmonics. The harmonic response is compared with the experimental values of Ref.~\cite{10.1063/5.0191184}. For the linear response, we employed Eq.~\ref{eq:P1}, using the experimental values for 5CB from Ref.~\cite{Vieweg:10}, i.e., $\alpha_\parallel=4.0$, and $\alpha_\perp= 3.2$. The third order response compares both the analytic third-harmonic generation expression (THG) of Eq.~\ref{eq:P3} and the N-th order general expression of Eqs.~\ref{eq:PN_x} and ~\ref{eq:PN_z}.} \label{fig:polar_plot}
\end{figure}

Consider now the linear optical response to an electric field which is linearly polarized in the plane containing the director and parameterized by angle $\varphi$ relative to the director, that is set along the $\hat{\vec{z}}$ direction. The electric field expressed as $\vec{E}(\theta) = E_0(\sin(\varphi),\cos(\varphi))$, impose a linear polarization in the polarization plane given by 
\begin{eqnarray}
 \vec{P}(\theta) &= &
 \begin{pmatrix}
 \langle\alpha\rangle_{\perp} & 0 \\
 0 &  \langle\alpha\rangle_{\parallel} 
\end{pmatrix}
 \begin{pmatrix}
E_0\sin(\varphi) \\
E_0\cos(\varphi) 
\end{pmatrix}\nonumber\\
&&= \alpha_0E_0
\begin{pmatrix}
\sin(\varphi) \\
\cos(\varphi) 
\end{pmatrix}
+\frac{\alpha_a S E_0 }{3}
\begin{pmatrix}
-\sin(\varphi) \\
2\cos(\varphi) 
\end{pmatrix}
\,.
\label{eq:P1}
\end{eqnarray}
Using such expression, one can calculate the polarization dependencies of the linear response for a nematic LC, assuming $\alpha_0$, $\alpha_a$, and $S$ are known. Or equivalently the order parameter, $S$, can be retrieved from anisotropy measurements of the LC, assuming that the optical properties of the constituent molecules are known. 
This is illustrated in Fig.~\ref{fig:polar_plot} using the parameters of the 5CB nematic LC. The first order polarizability with maximal yield along the nematic director, stands in clear contrast with the nonlinear components measured in Fig.~\ref{fig:1}. For direct comparison, however, one must proceed to calculate the nonlinear contributions.

\subsubsection{Generation of higher order harmonics} 
Proceeding to higher order processes, we directly consider the third-harmonic generation case, due to the vanishing second-order contribution in experiments~\cite{10.1063/5.0191184,experimental_paper}, arising from the inversion symmetry of the macroscopic molecular configurations. When imposing uniaxial symmetry considerations, the non-zero components of the third order nonlinear optical susceptibility $\chi^{(3)}$ tensor are the symmetry-axis component $\chi_{zzzz}$, the in-plane components $\chi_{xxxx}=\chi_{yyyy}$, $\chi_{xxyy}=\chi_{yyxx}=\chi_{xyxy}=\chi_{yxyx}=\chi_{xyyx}=\chi_{yxxy}$, and the mixed components $\chi_{zzxx}=\chi_{zzyy}$, $\chi_{xxzz}=\chi_{yyzz}=\chi_{xzxz}=\chi_{yzyz}$, and $\chi_{xzzx}=\chi_{yzzy}$. More generally, a susceptibility tensor of order $N$, must also fulfill intrinsic permutation symmetry and have an even number of transverse indices $T$, and a number of longitudinal components $L$ which relate by $T+L=N+1$~\cite{10.5555/1817101}. Here, longitudinal indices are defined along the director, while transverse indices denote directions perpendicular to it. The restriction to an even number of transverse indices follows from the cylindrical symmetry around the director and the inversion symmetry of the uniaxial distribution. 
We now impose a crude approximation in assuming the tensor to have a dominant component along the direction parallel to the molecular long-axis, and that all the other components are negligible. Such approximation is grounded on the optical response of the individual molecule typically being maximized along the molecular long-axis for the class of rod-like molecules, and as is also corroborated by first-principles simulations in later Fig.\ref{fig:hhg_gas}. This allows us to write the third-order susceptibility tensor of an individual molecule as
\begin{equation}
 \chi^{(3)}_{ijkl}(\vec{\nu}) = \chi_{\parallel}^{(3)}\nu_i\nu_j\nu_k\nu_l\,.
\end{equation}
With this, the averaged tensor reads
\begin{equation}
 \langle \chi^{(3)}_{ijkl}\rangle = \chi_{\parallel}^{(3)}\langle \nu_i\nu_j\nu_k\nu_l\rangle\,. \label{equivalent}
\end{equation}
Symmetry considerations require the averaged tensor to be structured as a linear combination of isotropic tensors and director dyads because, after the orientational average, the only remaining tensors compatible with uniaxial symmetry can be constructed from $\delta_{ij}$ and $n_i$
\begin{eqnarray}
 \langle \nu_i\nu_j\nu_k\nu_l\rangle &=& A_1(\delta_{ij}\delta_{kl} + \delta_{ik}\delta_{jl} + \delta_{il}\delta_{jk}) +  \nonumber\\
 && A_2(n_in_j\delta_{kl} +\ldots) + A_3 n_i n_j n_k n_l\,.
\end{eqnarray}
Here, $n_in_j\delta_{kl}+\ldots$ denotes the six distinct permutations $n_in_j\delta_{kl}$, $n_in_k\delta_{jl}$, $n_in_l\delta_{jk}$, $n_jn_k\delta_{il}$, $n_jn_l\delta_{ik}$, and $n_kn_l\delta_{ij}$.
The parameters $A_1$, $A_2$, and $A_3$ need to be determined, similarly to the linear optics case. The derivation of these parameters is outlined in Appendix (App.)~\ref{sec:derivation_A_coeff}, and it can be found that
\begin{eqnarray}
 A_1 &=& \frac{1}{15} - \frac{2}{21}S + \frac{1}{35}S_4 = \frac{7-10S+3S_4}{105}\,\nonumber\\
 A_2 &=& \frac{S-S_4}{7}\,\nonumber  \\
 A_3 &=& S_4,\nonumber
\end{eqnarray}
where we introduced the fourth-order scalar parameter $S_4 = \langle P_4 (\cos(\theta)) \rangle$.
From Eq.~\ref{equivalent} of the averaged third-order susceptibility, and the definition of the third-order polarization, $P^{(3)}_i = \sum_{jlk} \chi_{ijkl}^{(3)} E_jE_kE_l$, one can after some algebra, see App.~\ref{app:P3}, retrieve
\begin{align}
 \vec{P}^{(3)}&(\varphi) = 
 \nonumber \\
&\chi_{\parallel}^{(3)}E_0^3 \begin{pmatrix}
3[A_1 + A_2\cos^2(\varphi)] \sin(\varphi)  \\
[3(A_1 + A_2) + (3A_2 + A_3)\cos^2(\varphi)]\cos(\varphi) \nonumber
\end{pmatrix}.
\end{align}
To express everything in terms of the order parameter, $S$, we further approximate $\langle \cos(\theta)^4\rangle \approx \langle \cos^2(\theta)\rangle^2$, and arrive to a ``decoupling'' relation~\cite{KROGER200840}
\begin{equation}
 S_4 = \langle P_4(\cos(\theta))\rangle \approx P_4(\sqrt{\langle \cos^2(\theta)\rangle}),
\end{equation}
which grants 
\begin{equation}
 S_4 \approx \frac{-7 -10S + 35 S^2}{18}.
\end{equation}
Such expression can equivalently be obtained from the Schwarz inequality, see Ref.~\cite{degennes1993physics}. We note, however, that this closure does not recover the correct isotropic limit, for which $S=S_4=0$. A distribution-based closure, for example one based on a Maier-Saupe or Watson distribution, would therefore be preferable when the isotropic regime is explicitly considered.  
Using such approximation, we find 
 \begin{eqnarray}
 \vec{P}^{(3)}(\theta) &\approx& \chi_{\parallel}^{(3)}E_0^3 \times  \nonumber\\
&&
\begin{pmatrix}
\frac{1}{6}[(1-S)^2+ (1+4S-5S^2)\cos^2(\theta)] \sin(\theta)  \\
[\frac{1+S-2S^2}{3} + \frac{-2+S+10S^2}{9}\cos^2(\theta)]\cos(\theta) 
\end{pmatrix},\nonumber 
\label{eq:P3}
\end{eqnarray}
where again the order parameter $S$ can be spectroscopically retrieved from the anisotropy of the nonlinear response. The expression also allows for computing the polarization dependency of the third-order harmonic response from an ensemble of uniaxial molecules with an order parameter $S$, for which the nonlinear response of each individual molecule is dominated by the susceptibility along the molecular axis.
The $3^\mathrm{rd}$ harmonic yield as a function of polarization axis is shown in Fig.~\ref{fig:polar_plot}. It displays a dominant response along the director, in complete disagreement with the experimental result of Ref.~\cite{10.1063/5.0191184}. The reason for this discrepancy will be examined later, and involves relaxing the approximation of having independent molecular emitters.

The same approach can be extended to the higher order case, as discussed in App.~\ref{sec:higher-order}. For arbitrary order, however, we introduce an additional mean-angle closure rather than evaluating all orientational moments exactly. In this closure, the molecular axis is represented by an effective longitudinal projection $\mu$ and transverse projection $C$, with $\mu^2=(1+2S)/3$ and $C=\sqrt{1-\mu^2}$. Increasing the order parameter therefore increases the projection along the director and suppresses the effective transverse component.
Applying these expressions to the nonlinear response for harmonic orders $3$ to $7$, we obtain the results shown in Fig.~\ref{fig:polar_plot}. For the $3^\mathrm{rd}$ order, the $N^\mathrm{th}$-order expression reproduces perfectly the analytical expression of Eq.~\ref{eq:P3}. However, using such perturbative nonlinear optics approaches, we still cannot reproduce the experimental observation of Ref.~\cite{10.1063/5.0191184}, provided here for the E7 liquid crystal mixture, which predominantly consist of the 5CB liquid crystal. This shows that the disagreement is not related to the perturbative order, but rather to the approximation of an independent-emitter picture. Instead, one has to relax the assumption that the nonlinear response is dominated by isolated molecular axes, for instance by allowing intermolecular hybridization to generate effective response channels perpendicular to the director. We conclude that the experimental observation of the nonlinear response cannot be explained as an ensemble average of the isolated molecular responses, a fact which indicates that nonlinear spectroscopy of LCs is intrinsically sensitive to the hybridization between the molecules and thus is formed from a more collective rather than independent microscopic response. 

\subsection{Semi-empirical tight-binding approach} \label{sec:2}
\begin{figure} 
	\includegraphics[width=0.6\columnwidth]{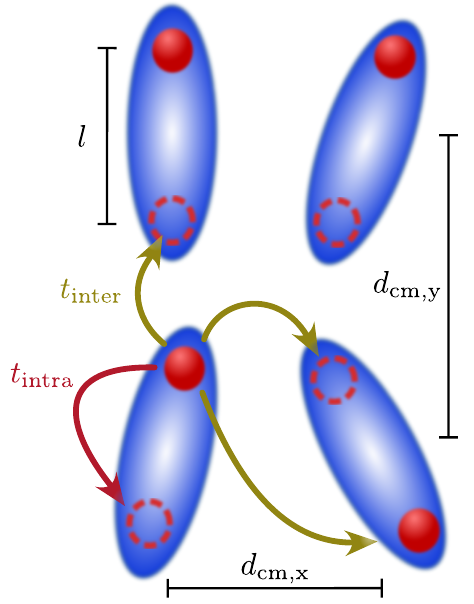}
	\caption{Illustration of the tight-binding model with intramolecular $t_\text{intra}$ and intermolecular $t_\text{inter}$ couplings illustrated by red and yellow arrows. Further, the intermolecular center-of-mass distances $d_{\text{cm,x}}$, $d_{\text{cm,y}}$ and molecular length $l$ are illustrated.} \label{fig:tight_binding_illustration}
\end{figure}
To move beyond the conventional perturbative expansions of nonlinear optics, we now turn to consider microscopic time-dependent quantum simulations. The underlying structural properties and electron processes that govern light-driven soft matter systems, such as across different LC mesophases, can be elucidated considering a 2D tight-binding model in the laser polarization plane. Such an approach has proven fruitful for elucidating the polarization dependencies and the role of intra- and intermolecular couplings for harmonic generation in the related class of molecular crystal materials~\cite{Wiechmann2025,Bielke2025}. The flexibility of tight-binding modeling techniques allows us to include a semi-empirical structural description to advance from the scope of crystalline matter, towards elucidating the wider range of soft matter systems. While the underlying model parameters for the present case rely on accurate first-principles calculations and experimentally extracted properties regarding a specific LC, the success of the simplistic and generic model provides confidence that the observed signatures and results translate to a wider range of soft-matter systems. 

The molecules forming the LC structure are, to a large degree, defined by a few key structural properties: the molecules are elongated with a rigid molecular backbone that defines the long molecular axis, and the extremities of the molecules typically consist of less influential functional groups~\cite{RevModPhys.46.617}. Knowledge of these defining properties allows us to simplify the theoretical description of the structure of the LC using a tight-binding approach in which the molecules of the LC are modeled as individual two-site systems, as illustrated by Fig.~\ref{fig:tight_binding_illustration}. These two sites are separated by the molecular length $l$, which represents the rigid elongated part of a non-polar molecule and sets its long axis. The two sites of the simplified molecular model are coupled by an intramolecular coupling $t_\text{intra}$ that gives rise to the energy gap between the highest occupied and lowest unoccupied electron energy state of the molecule. By including the intramolecular and intermolecular couplings, the tight-binding model grants a much more realistic modeling of the LCs, compated to the previous section, as it allows for capturing the key structural characteristics and couplings, while retaining the simplicity of a model.

While the underlying assumption of molecular rigidity of the model can seem like a crude approximation, we are only interested here in the ultrafast electron dynamics occurring on the few-femtosecond timescale, during the interaction between the LC and the femtosecond laser pulse. The fact that molecules may be flexible, or exhibit thermal and laser-induced structural fluctuations or phase transitions, are phenomena occurring on a much longer timescale than the ultrafast electronic response. Hence, the influence of such structural dynamics on the electron motion is neglected here. Since interactions between molecules are small, we assume a tight-binding interaction, and only nearest-neighbor intermolecular interactions are included, in the form of an intermolecular coupling $t_\text{inter}$ parameter. This leads to the following tight-binding Hamiltonian in atomic units
\begin{align}
	H =  &\sum_{\lbrace i,j \rbrace \in S_\text{intra}}  \left(-t_\text{intra}  c^\dagger_i c_j +  \text{h.c.}\right)   \nonumber \\+ &\sum_{\lbrace n,m \rbrace \in S_\text{inter}}  \left(-t_\text{inter}^{nm}  c^\dagger_n c_m + \text{h.c.}\right), \nonumber
\end{align}
where $S_\text{intra}$ ($S_\text{inter}$) is the set of intramolecular (intermolecular) couplings and h.c.~denotes the hermitian conjugate. The major difference between LCs and molecular crystals lies in the order parameter, which determines the molecular orientations and intermolecular distances. To capture this key effect in our model, the intermolecular interactions are parameterized for different intersite distances $d$ with Harrison's law $t_\text{inter}^{mn} = t_{\text{inter},0}\left(d_0/d^{mn}\right)^2$ expanded around $d_0$ and $t_{\text{inter},0}$~\cite{PhysRevB.10.1516}. While the dipole moments of the molecules are critical for the structural properties of the molecules, we are only concerned in the present work with the electron dynamics on the femtosecond timescale, and the effect of the dipole-dipole interaction on the molecular orientation and its fluctuation in time is thus neglected. This implies that our model considers a collection of uniaxial rod-like molecules, each with a center of symmetry. Ferroelectric effects that break inversion symmetry, and would translate into second-harmonic generation and higher-order even harmonics, will thus not be captured in the current model. This assumption is justified by HHG experiments on nematic LCs~\cite{10.1063/5.0191184, experimental_paper} showing no signature of even harmonics, obtained using laser spot sizes across a macroscopic domain of molecules. The proposed model should therefore still be able to shed light on the interplay between the order parameter, the anisotropy in the LC, and the corresponding HHG spectra.

\subsubsection{First-principles molecular parameters} \label{sec:th}

\begin{figure} 
	\includegraphics[width=\columnwidth]{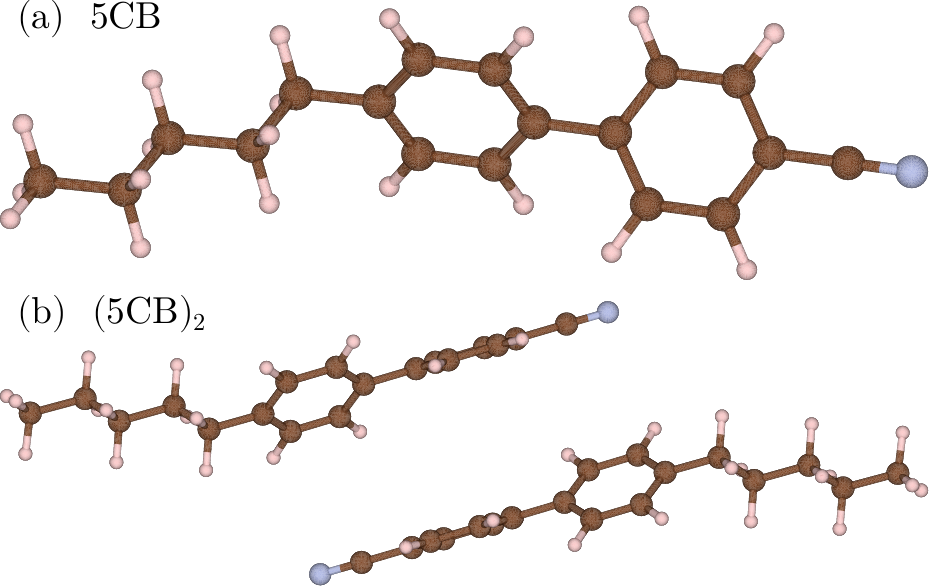}
	\caption{Lowest energy structural geometry obtained by geometry optimization for (a) the isolated 5CB molecule and (b) the 5CB dimer, which exhibits antiparallel dimerization.} \label{fig:geometry_optim}
\end{figure}

For a realistic description of the electronic structure of LCs, we perform first-principles geometry optimization and electronic structure theory using the Octopus software~\cite{10.1063/1.5142502}, taking 4'-pentyl[1,1'-biphenyl]-4-carbonitril (5CB) as a prototypical example, for which HHG can be measured experimentally~\cite{10.1063/5.0191184}. Structural geometry optimization is performed utilizing the FIRE algorithm~\cite{PhysRevLett.97.170201} within a real-space formulation of density-functional theory~\cite{PhysRev.136.B864} utilizing the PBE functional~\cite{PhysRevLett.77.3865} and including van der Waals interactions~\cite{10.1063/1.3382344}. The single 5CB molecule and configurations of 5CB dimers are considered with initial parameters retrieved from Refs.~\cite{Hanemann01111995,10.1063/1.1494799,C4CP04933K}. Convergence is obtained with a grid spacing of $0.2$ a.u. granting a molecular structure of 5CB as depicted in Fig.~\ref{fig:geometry_optim}~(a) and provided in Tab.~\ref{tab:coord} of App.\ref{App:P4}. For the molecular dimer, we obtained the lowest energy configuration to be the antiparallel dimer configuration as depicted in Fig.~\ref{fig:geometry_optim}~(b), as also found in Ref.~\cite{C4CP04933K}.

\begin{figure} 
	\includegraphics[width=\columnwidth]{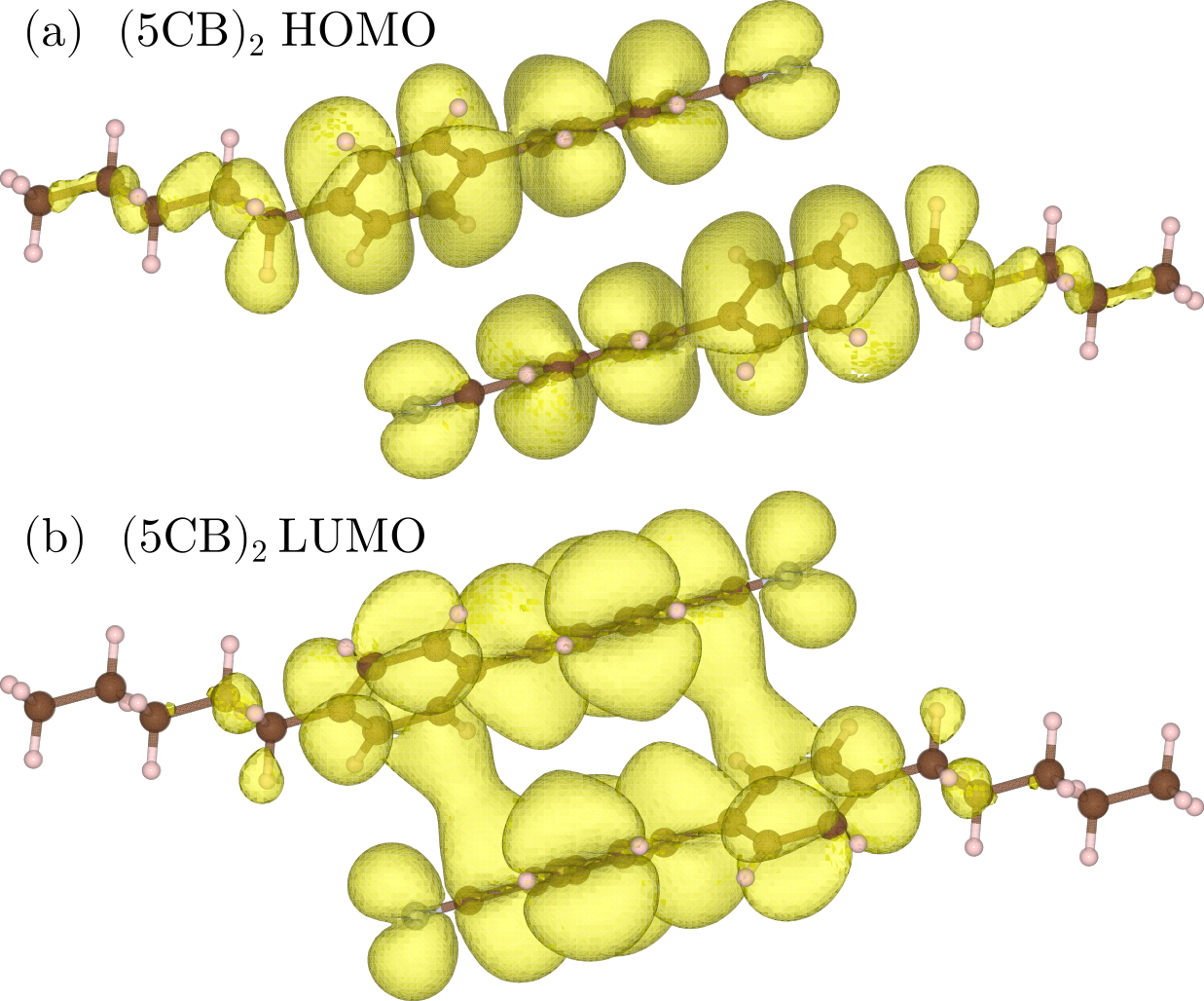}
	\caption{(a) Highest occupied molecular orbital (HOMO) and (b) lowest unoccupied molecular orbital (LUMO) of the antiparallel 5CB dimer.} \label{fig:homo_lumo}
\end{figure}

Accurate electronic structure calculations are hereafter achieved utilizing the B3LYP hybrid functional~\cite{10.1063/1.464913} at a converged grid spacing of $0.25$ a.u. We find a gap between the HOMO and the LUMO of the 5CB molecule to be $4.61$ eV, consistent with literature values~\cite{Sharma2021,UPADHYAY201541} which grants $t_\text{intra} =  0.085$. For the antiparallel 5CB dimer, we obtain the HOMO and LUMO wavefunctions depicted in Fig.~\ref{fig:homo_lumo}, showing a clear hybridization between molecules of the dimer. The dimer HOMO-LUMO gap is obtained to $4.43$ eV, allowing us to set $t_{\text{inter}_0} =  0.048 t_\text{intra}$. The electron binding energy for the molecule and dimer is obtained to be $6.60$ eV and $6.45$ eV, respectively. We define the molecular length $l = 9.75$ \AA~to be that of the rigid molecular part that contributes to dimerization, excluding the flexible pentyl groups. The equilibrium intermolecular distance between the molecular ends is obtained to be $d_0 = 4.78$ \AA~for the converged lowest energy antiparallel dimer configuration.

\subsubsection{Semi-empirical structural representation}
\begin{figure} 
	\includegraphics[width=\columnwidth]{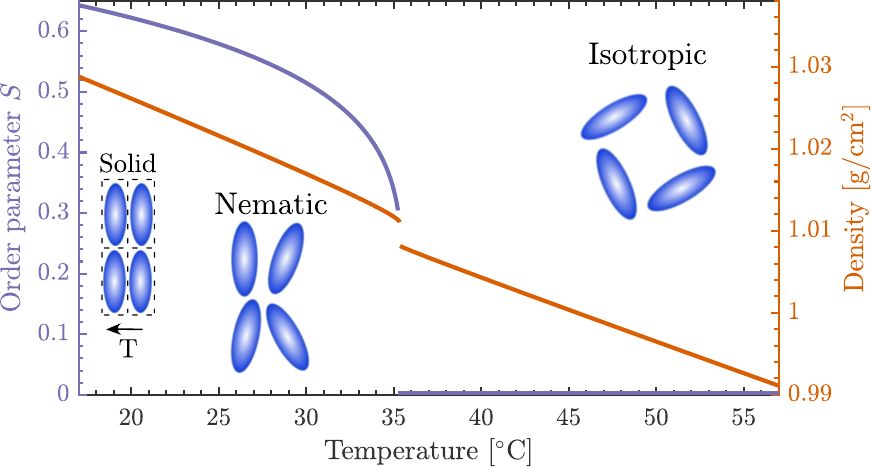}
	\caption{Order parameter and density retrieved as a function of temperature from Refs.~\cite{doi:10.1021/acs.jpclett.1c03517,Magnuson01121995,Roushdy01112006,doi:10.1021/jp711211w,https://doi.org/10.1002/crat.200900162}
    with illustrations of corresponding LC mesophases.} \label{fig:order_density_temperature}
\end{figure}

Moving towards a simple but accurate representation of the LC structure, we consider two-dimensional simulations corresponding to the laser-polarization plane within a LC, wherein we assume the nematic director is contained. The two-dimensional approach is perfectly suited to describe a realistic liquid flat jet experimental setup, for which flow-induced alignment assures the director to be aligned in the plane of the flat jet. Measuring in transmission geometry, thus offers a direct probe of the director in the polarization plane of the laser. To simulate the change of structural order between LC mesophases, we account for the LC order parameter, which is given in 2D as   
\begin{equation}
S = \frac{1}{N} \sum_{i=1,2,3,...}^N   \left( 2\cos(\theta_i)^2 -1 \right)\,,    \label{eq:order}
\end{equation}
with $\theta_i$ being the angle between each molecular unit vector $\bm{e}_i$, and $\bm n$ denoting the nematic director, defined in practice as the principal axis of the corresponding two-dimensional alignment tensor. For the isotropic phase, the order parameter attains $S=0$, and for the nematic phase, it attains $0<S<1$, where the limiting hypothetical case of $S=1$ corresponds to perfectly aligned molecules. Our model considers a nematic director along the y-direction and parametrizes the order parameter $S$ from its temperature dependence as prescribed from 
experiments~\cite{doi:10.1021/acs.jpclett.1c03517,Magnuson01121995,Roushdy01112006}
and depicted in Fig.~\ref{fig:order_density_temperature}, giving e.g.,  
$S = 0.52$ ($S = 0$) for the nematic (isotropic) phase at $T = 30^\circ$C ($T = 50^\circ$C). 
The order parameter is obtained by sampling the molecular rotations around the nematic director with a head-tail-symmetric von Mises distribution for the variable $x=2\theta$, $p(x)=\exp\left[\kappa \cos\left(x\right)\right]/(2\pi I_0(\kappa))$. Adjusting the associated measure of concentration $\kappa$ sets an order parameter, which in the macroscopic limit is given by $S_{\text{macro}} = I_1(\kappa)/I_0(\kappa)$ with $I_n$ being the modified Bessel function of the first kind. The inversion symmetry of each molecule allows for improving the computational efficiency by sampling a symmetry-reduced von Mises distribution to neglect the longest-distance intermolecular couplings, compared to their dominant short-distance counterparts. 

\begin{figure} 
	\includegraphics[width=\columnwidth]{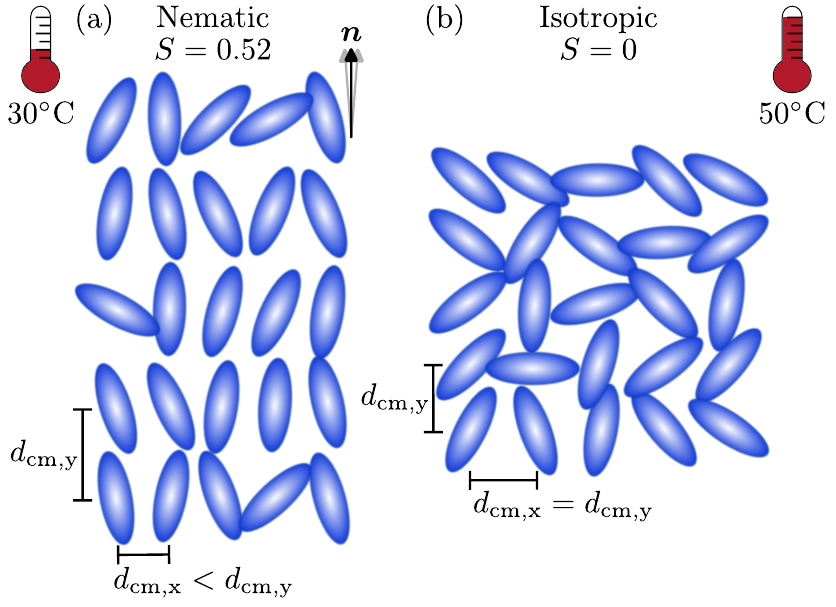}
	\caption{Illustration of the tight-binding model's structural dependency with respect to the change of density and order parameter caused by temperature variations retrieved from Fig.~\ref{fig:order_density_temperature}. For (a) the low temperature nematic phase, molecules tend to pack more closely along the direction perpendicular to the director $\bm n$, while expanding along the director, causing intermolecular distances to relate as $d_{\text{cm,x}} < d_{\text{cm,y}}$. For the high temperature isotropic phase of (b), the vanishing order parameter causes equidistant molecular distributions $d_{\text{cm,x}} \approx d_{\text{cm,y}}$, with an overall increasing molecular occupation volume.} \label{fig:sketch_structure}
\end{figure}

We find that, in addition to the distribution of molecular orientations, the change of microscopic molecular packing also contributes significantly to describing the spectroscopic signatures of the LC. To account for such effects, we propose a minimal structural closure of the center-of-mass intermolecular distances of $d_{\text{cm,y}} = lS + d_{\text{cm,x}}$, illustrated in Fig.~\ref{fig:sketch_structure}. This relation is an ansatz calibrated by the temperature-dependent density and radial-distribution data, rather than a general consequence of nematic order. Such packing ensures equal distances between the sites of neighboring molecules on average, granting e.g. identical center-of-mass intermolecular distances $d_{\text{cm,x}} = d_{\text{cm,y}}$ for the isotropic phase $S = 0$, and equidistant spacing between intermolecular sites $d_{\text{cm,y}} = l + d_{\text{cm,x}}$ for the aligned phase of $S = 1$. The volume associated with each molecule is obtained as $V_s = d_{\text{cm,y}} d_{\text{cm,x}}^2$, and determined from extracted temperature-dependent density measurements of Refs.~\cite{doi:10.1021/acs.jpclett.1c03517,doi:10.1021/jp711211w,https://doi.org/10.1002/crat.200900162}, as depicted in Fig.~\ref{fig:order_density_temperature}, with molecular weight of $249.34$ g/mol~\cite{Hanemann01111995}. We compensate for the geometric consequences of the reduced dimensionality of the model by adjusting the simulation volume $V_s$, compared to the actual volume $V$, with $V_s = 3\sqrt{3}V/4$. This accounts for the decrease in packing efficiency within our disordered square cuboid 2D model compared to a true 3D body-centered-cubic crystal, as is the typical starting point for LC molecular dynamics simulations~\cite{10.1063/1.1482702}. 

To accurately describe the macroscopic structural and optical response of the LC, we sample across numerous molecular configuration snapshots with periodic boundary conditions, each containing $N=100$ molecules.
The number of molecules within the simulation box is determined from convergence studies while verifying that the electron quiver radius is much less than the simulation box with the given laser parameters. 

 \begin{figure} 
	\includegraphics[width=\columnwidth]{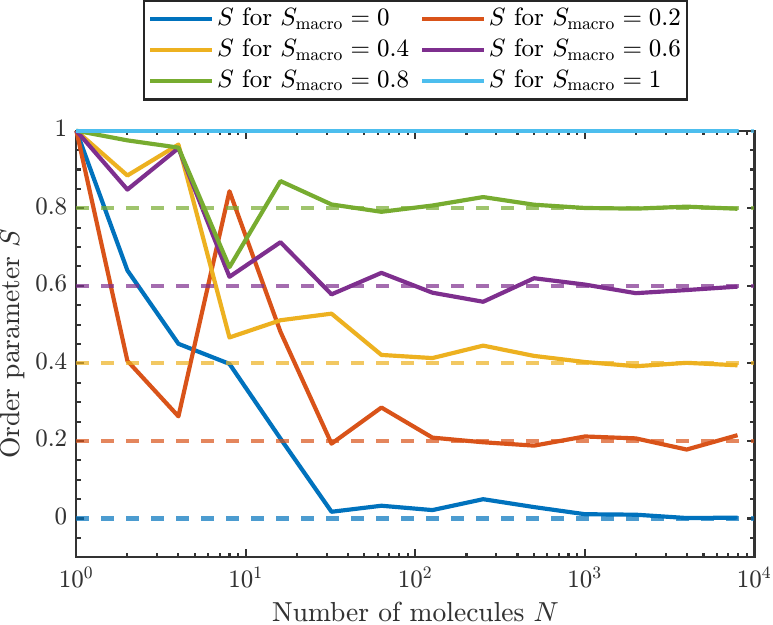}
	\caption{Retrieved order parameter $S$ of Eq.~\eqref{eq:order} as a function of the number of molecules within the simulation box. The macroscopic limits are depicted with dashed curves.} \label{fig:convergence_order}
\end{figure}

\begin{figure} 
	\includegraphics[width=\columnwidth]{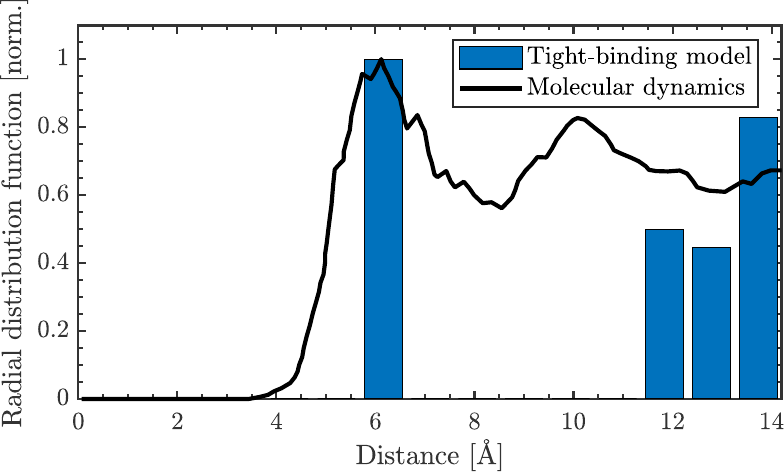}
	\caption{Center of mass radial distribution function for 5CB molecules in the nematic LC at a temperature of $26.85 ^\circ$C for (a) the tight-binding model and (b) molecular dynamics simulations of Ref.~\cite{10.1063/1.1482702}.} \label{fig:radial_dist}
\end{figure}
 
We further verify in Fig.~\ref{fig:convergence_order} that upon accumulating multiple snapshots of the LC structure, the obtained order parameter converges to its macroscopic limit. Similarly, coherent addition of the light-induced current from such molecular snapshots converges to the macroscopic optical response, as previously explored for HHG in liquids~\cite{doi:10.1021/acs.jctc.2c00235,Mondal2023}. However, since the optical signatures exhibit a slower convergence than the order parameter, in the following results, a minimum of $10^4$ molecules is considered for each parameter setting. As a general trend, systems of higher order parameter tend to converge faster towards their macroscopic limit when increasing the number of molecules, while isotropic systems tend to require a higher degree of sampling to retrieve an accurate representation of their structural disorder. 
Accuracy of the simplified LC structural description with respect to the structural variations caused by the temperature-dependent density changes of Fig.~\ref{fig:order_density_temperature} is verified by comparing the obtained radial distribution function with more elaborate molecular dynamics simulations of Ref.~\cite{10.1063/1.1482702} in Fig.~\ref{fig:radial_dist}. As observed, the nearest-neighbor center-of-mass distance of $d_{\text{cm,x}} = 6.1$ \AA~for our tight-binding model at a temperature of $26.85 ^\circ$C fits exactly with the retrieved maximum of the radial distribution from the molecular dynamics simulations. We note that the addition of a random displacement within $0.075 l$ of the molecular center of mass position will broaden the radial distribution function of the model, to provide a similar peak width as molecular dynamics simulations of Ref.~\cite{10.1063/1.1482702}. In the present work, we do not consider such additional statistical disorder, but will include such an improved structural description for the experimental investigation of Ref.~\cite{experimental_paper}. We acknowledge that due to the reduced dimensionality of the simplified model, next-to-nearest-neighbor distances are overestimated compared to molecular dynamics simulations. Motivated by obtaining a simple physical description, we neglect the associated next-to-nearest-neighbor couplings, which would require a full three-dimensional model to grant more accurate structural properties. 

\section{High-order harmonic spectroscopy} \label{SectionHHG}
\begin{figure} 
	\includegraphics[width=\columnwidth]{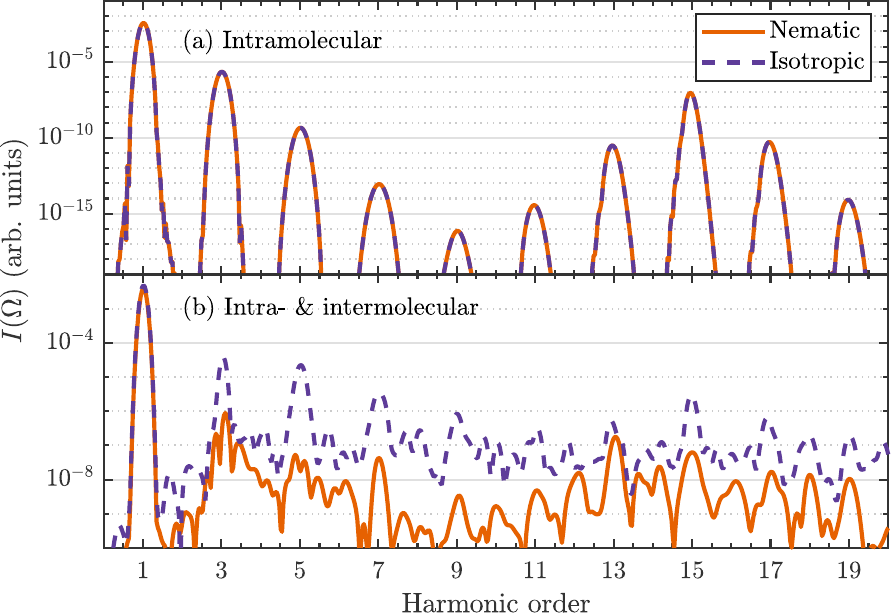}
	\caption{Orientational averaged high-order harmonic spectra generated from the nematic and isotropic 5CB LC phase by a $0.1$ TW/cm$^2$ sin$^2$ pulse of $80$ fs FWHM and $\lambda = 4000$ nm. Considering (a) independent molecular emitters containing only intramolecular electron dynamics and (b) interacting molecules exhibiting both intra- and intermolecular electron dynamics. 
    } \label{fig:hhg_spec}
\end{figure}

For time-dependent calculations both intra- and intermolecular hoppings are subject to Peierls substitution, $t_{l,j} \rightarrow t_{l,j}\exp[-i\bm{A}(t)\cdot(\bm{r}_l-\bm{r}_j)]$, with intersite distance vector $\bm{r}_{l,j} = \bm{r}_l - \bm{r}_j$ and electromagnetic vector-potential $\bm{A}(t)$~\cite{Peierls1933}. We consider an $80$ fs FWHM sin$^2$ electromagnetic field with wavelength $\lambda = 4000$ nm and peak intensity of $0.1$ TW/cm$^2$. Propagation is performed with a Runge-Kutta (4,5) propagator, utilizing a converged timestep of $1$ a.u. upon which the time-dependent current $\bm j(t)$ is evaluated with the procedure of Ref.~\cite{PhysRevB.84.085128}. The harmonic spectrum is obtained from Larmor's formula $I(\Omega) \propto \Omega^2 \norm{\bm j(\Omega)}^2$, applying a $\cos^8$ window function to reduce numerical artifacts from the Fourier transform~\cite{jackson_classical_1999}.  
Initially, we consider the orientational averaged HHG spectra for the nematic and isotropic LC phases in Fig.~\ref{fig:hhg_spec}. Considering only intramolecular interactions in Fig.~\ref{fig:hhg_spec}~(a), identical harmonic signatures are observed across both phases, as is expected for orientational averaging of two distributions of identical independent emitters. 
When comparing total experimental spectral yields, one should keep in mind that the temperature-dependent density of the LC will cause the number of emitters within a laser focal spot volume to vary between different phases. While Fig.~\ref{fig:hhg_spec} compares an equal number of emitters, the laser focus volume will contain $\sim 2 \%$ more molecules in the nematic, compared to the isotropic case, an effect likely to cause an enhancement of retrieved harmonic yield. Such enhancement is observed in experiments considering the E7 LC mixture~\cite{10.1063/5.0191184}, however, in vastly superior magnitude, which also varies for different spectral regimes. A varying enhancement across different frequency regimes is a property that cannot arise from the statistical ordering of independent emitters alone, indicating that a more complex underlying physical process is at play.

When adding intermolecular interactions in Fig.~\ref{fig:hhg_spec}~(b), both phases exhibit a nonlinear optical response which is significantly enhanced in magnitude compared to Fig.~\ref{fig:hhg_spec}~(a).
Additionally, the vastly different spectroscopic response of the two phases suggests that the nonlinear intermolecular electron dynamics enables a sensitivity towards the underlying structural properties. In doing so, intermolecular electron dynamics also provides a possible mechanism, which can cause a varying enhancement across different frequency regimes in experiments~\cite{10.1063/5.0191184}. However, contradictory to the experiments in the E7 LC mixture, we find for 5CB that the harmonic signal is decreased rather than increased in the nematic phase, compared to the liquid phase. We attribute the likely cause of this discrepancy to macroscopic effects such as phase matching, which is expected to be significantly altered by the different structure and density between the two phases. Whether an explanation could also be found in the more multifaceted intermolecular couplings of a LC mixture, or enabled by a more evolved modeling approach, is a topic for further investigations. We, however, note that liquid jet experiments with 5CB seem to corroborate the simulations at hand, granting a higher nonlinear component for the isotropic phase, compared to the nematic phase~\cite{experimental_paper}.
Since the intermolecular coupling alone is not an easily assessible control knob in experiments, to achieve a more in-depth interpretation of the underlying mechanisms, we look for additional spectroscopic information in the anisotropic features of the harmonic response, while noting from Fig.~\ref{fig:hhg_spec} that high-frequency harmonics in general are very sensitive to the intermolecular couplings.

\section{Polarization-resolved spectroscopy}  \label{Section3}
The ability in experiments to vary the driving laser polarization with respect to the target structure opens another dimension of spectroscopic information. The gathered spectroscopic insights and underlying physical interpretations from such an investigation have historically depended on the phase of matter at hand. To provide a rigorous investigation across such different phases, we consider high-harmonic polarization spectroscopy initially from the well-established gaseous and crystalline phases, to later explore the intermediary soft matter and liquid phases. In order to compare and contrast the anisotropic spectroscopic features on equal footing, we consider the 5CB molecule across all such phases, while employing state-of-the-art electronic structure and dynamics methods.   

\subsection{Gas phase}
For gas phase HHG, one observes an isotropic spectral response with constant harmonic yield for all polarization directions, if driving a macroscopic gaseous collection of molecules that are randomly oriented. Thus, typically no additional spectroscopic information can be retrieved intuitively by a polarization scan, unless some symmetry-breaking property comes into play, such as chirality or magnetic field effects, alike circular dichroism~\cite{Cireasa2015,Ayuso:22}. If, however, one employs techniques for orienting and aligning molecules with light~\cite{RevModPhys.75.543}, additional spectroscopic information can be gathered, enabled by the anistropic structural sensitivity of the recolliding electron, a feature which has been utilized for imaging molecular orbitals~\cite{Itatani2004,Vozzi2011}, or attosecond nuclear and electron motion~\cite{doi:10.1126/science.1123904,doi:10.1126/science.aab2160}. As a general trend, the harmonic yield is maximized (minimized) when the laser polarization direction is parallel (perpendicular) to the molecular long axis. A feature that can be expected from the trend observed for independent aligned emitters of Fig.~\ref{fig:polar_plot}. Contradictory trends are observed for specific species and laser parameters, while still attributed to single-molecular response, modulated by characteristic structural interference anisotropies~\cite{Kanai2005,PhysRevLett.95.153902,PhysRevLett.88.183903}. To investigate the spectroscopic response from gas-phase 5CB molecules, we consider identical laser parameters as of Fig.~\ref{fig:hhg_spec}, with molecular coordinates of Fig.~\ref{fig:geometry_optim} applying
self-interaction-corrected PBE TDDFT~\cite{CLegrand_2002,PhysRevLett.77.3865} within the Octopus simulation package~\cite{10.1063/1.5142502}. Converged spectroscopic response is retrieved utilizing a grid spacing of $0.2$ a.u. and a timestep of $0.5$ a.u. for a spherical simulation box of 45 a.u. radius. 

\begin{figure} 
	\includegraphics[width=\columnwidth]{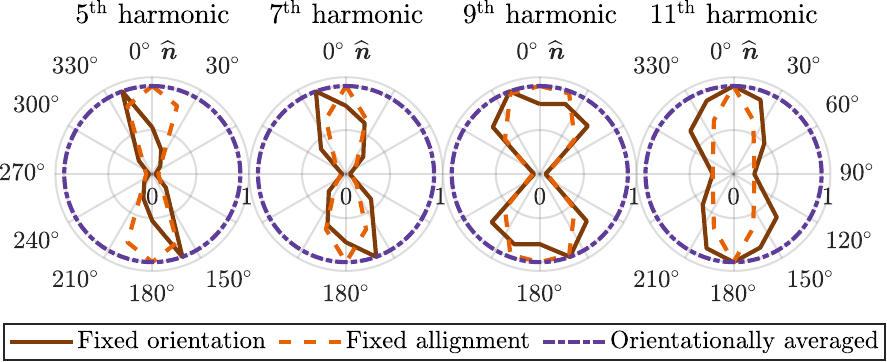}
	\caption{Harmonic intensity as a function of polarization direction considering a 5CB molecular gas being oriented, aligned, or randomly distributed with the principal molecular axis along $\theta = 0^\circ$. The harmonic yield is normalized for each phase, with field-inversion symmetry imposed on the polarization response, and utilizing identical laser parameters to Fig.~\ref{fig:hhg_spec}.} \label{fig:hhg_gas}
\end{figure}

Well beyond the $13$ a.u. estimated electron excursion amplitude, a complex absorbing boundary potential of $-1$ a.u.~height and $15$ a.u.~width is utilized to minimize reflection errors in the $\sim$ eV frequency regime~\cite{DeGiovannini2015}. Simulations are performed by sampling $6$ molecular orientations symmetrized with the long molecular axis in the polarization plane. 
In Fig.~\ref{fig:hhg_gas}, the harmonic yield is provided considering both the orientational averaged signal from a randomly distributed molecular gas, the coherent response of a collection of completely orientationally fixed molecules, and the coherent response of an aligned molecular gas with the laser polarization defined parallel or antiparallel to the molecular principal axis of inertia. As expected, randomly aligned molecules represented by orientational averaging grant an isotropic spectroscopic response, while the oriented and aligned molecular gas grant a maximum harmonic yield when driving along the long molecular axis, with a reduced yield along perpendicular directions, consistent with literature findings~\cite{Kanai2005,PhysRevLett.95.153902,PhysRevLett.88.183903}. 

Since the gas phase is associated with large intermolecular distances and a general neglect of intermolecular interactions, the features of Fig.~\ref{fig:hhg_gas} represent a limiting case for comparison, in which intermolecular interactions are neglected, allowing for the use of evolved TDDFT calculations. To get a perspective of the other limiting case, where symmetry considerations also permit large-scale TDDFT calculations, we turn to consider the crystalline phase.

\subsection{Crystalline phase}
\begin{figure} 
	\includegraphics[width=\columnwidth]{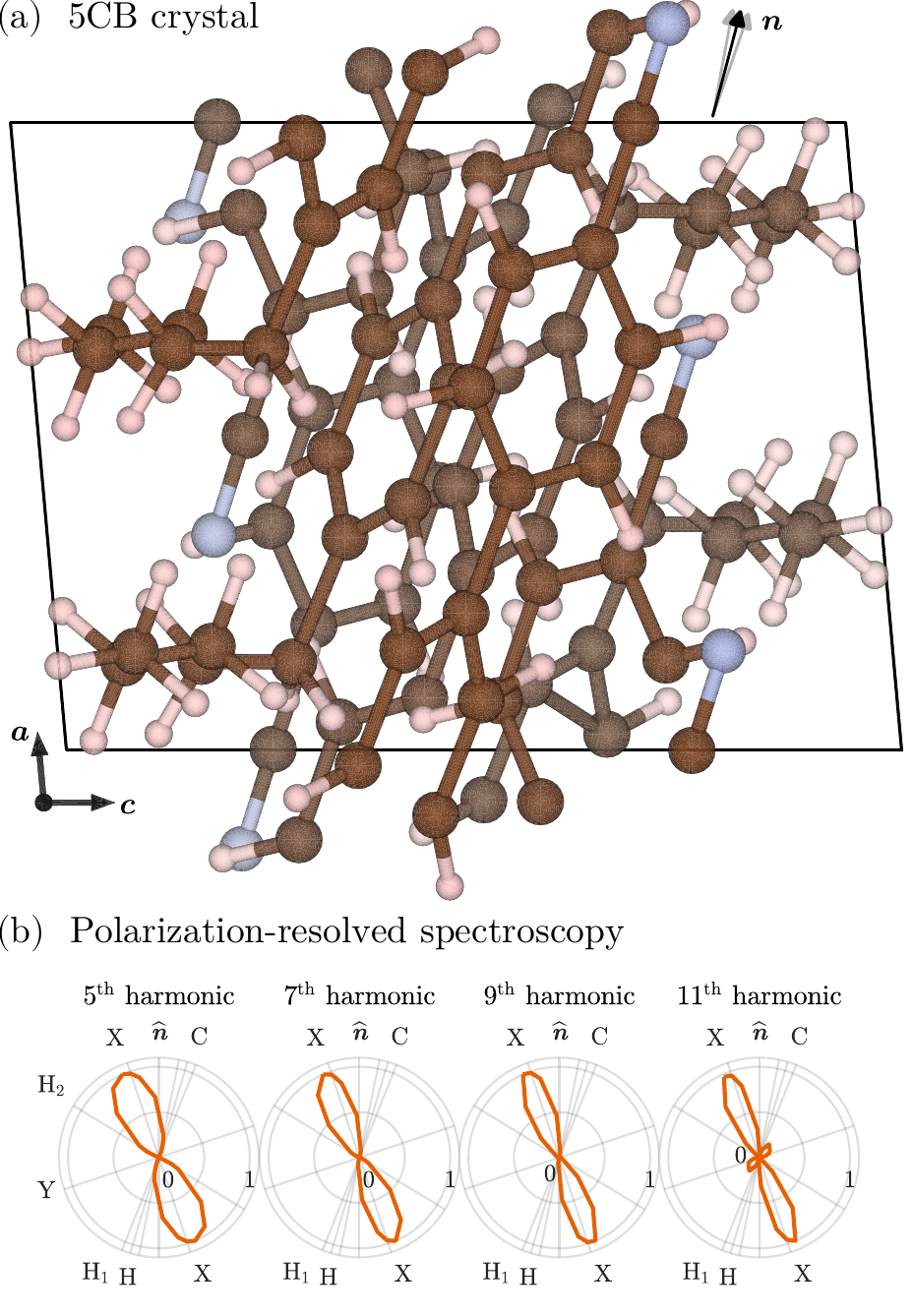}
	\caption{(a) 5CB crystal structure projected along $[0, 1, 0]$ with lattice vectors $\bm a$ and $\bm c$ given alongside main molecular axis direction $\bm n$. (b) Normalized harmonic yield with respect to laser-polarization direction in the $[0, 1, 0]$-plane imposing field-inversion symmetry and laser parameters of Fig.~\ref{fig:hhg_spec}.} \label{fig:hhg_crystal}
\end{figure}
In the crystalline phase, varying the crystallographic orientation with respect to the driving field has been widely utilized to probe symmetries of the crystal structure~\cite{Ghimire2011,Schubert2014}, associated band structure couplings and curvatures~\cite{PhysRevLett.120.243903,PhysRevA.96.063412} and identify exciton responses~\cite{Langer2016}. For consistency, we similarly consider the HHG spectroscopic response of the crystalline phase of the 5CB LC. The crystalline phase is formed at the lowest temperature regime, where it exhibits a monoclinic geometry with crystal vectors $a = 8.25$ \AA, $ b = 8.25$ \AA, $c = 10.94$ \AA, and angle $\beta = 95.093 ^\circ$ ~\cite{Hanemann01111995}. We insert hydrogen atoms into the literature structure and perform symmetry-conserving FIRE geometry optimization within van der Waals corrected PBE DFT utilizing a grid spacing of $0.16$ a.u. to obtain a converged band gap energy of $2.66$ eV~\cite{PhysRevLett.97.170201,10.1063/1.3382344,PhysRevLett.77.3865}. In comparison, a $\Gamma$-point calculation utilizing the B3LYP hybrid grants a corrected band gap of $3.54$ eV~\cite{10.1063/1.464913}. The time-dependent calculations for obtaining the optical anisotropy converge utilizing a $2 \times 2 \times 2 $ Monkhorst-Pack grid~\cite{PhysRevB.13.5188}, a $0.3$ a.u. spatial, and $0.1$ a.u. temporal grid spacing. The unit cell contains $4$ molecules, which are packed closely to form a structural arrangement of bilayers, with two parallel molecules followed by two antiparallel molecules that are separated from neighboring configurations by out-of-plane pentyl chains, see Fig.~\ref{fig:hhg_crystal}~(a). Since the molecules do not attain identical long molecular axes in 3D, a direct comparison to the gas phase properties is difficult to perform on equal footing. However, as a result of the crystal symmetry, the average molecular alignment is confined well in the $[0,1,0]$-plane of Fig.~\ref{fig:hhg_crystal}~(a), to have an average molecular axis tilted $23.5 ^\circ$ from $\bm a$. Considering the harmonic yield with respect to the polarization of the driving field in the $[0,1,0]$-plane in Fig.~\ref{fig:hhg_crystal}~(b), a clear maximum of harmonic yield is retrieved along the $\overline{\Gamma X}$ direction as electrons are driven along the layered structure, a direction rotated with $18.5 ^\circ$ from the average long molecular axis. The crystalline phase thus showcases, similar to literature~\cite{Ghimire2011,Schubert2014,PhysRevLett.120.243903,PhysRevA.96.063412,Langer2016},  that symmetries of the crystal structure are governing the harmonic emission and no longer the alignment of the long molecular axis. In its essence, these properties arise from the maximally delocalized electronic Bloch states of the crystal, which arise from the maximally hybridized molecular orbitals formed by the strong intermolecular coupling.  

Since the range of liquid and soft matter materials can explore structural regimes between the gas and crystalline phases, we expect the spectroscopic response to showcase the gradual change between such underlying gaseous single-molecular and crystalline many-molecular processes. For phases resembling a gas, the spectroscopic response is expected to originate from individual molecular processes that can be subject to averaging across statistical structural distributions. On the other hand, phases closer to the crystalline phase are expected to exhibit an intrinsically collective many-molecular response arising from delocalized intermolecular electron dynamics. 
From the insights obtained in the total harmonic spectrum of Fig.~\ref{fig:hhg_spec}, we see strong indications that intermolecular interactions already contribute in the liquid and nematic phases, hinting at a possible interpretation that should be found from delocalized electron dynamics across multiple molecules. As will arise from the following section, the anisotropies of the nematic LC nonlinear response will corroborate this intermolecular electronic picture.

\subsection{Liquid crystal phases}
\begin{figure} 
	\includegraphics[width=\columnwidth]{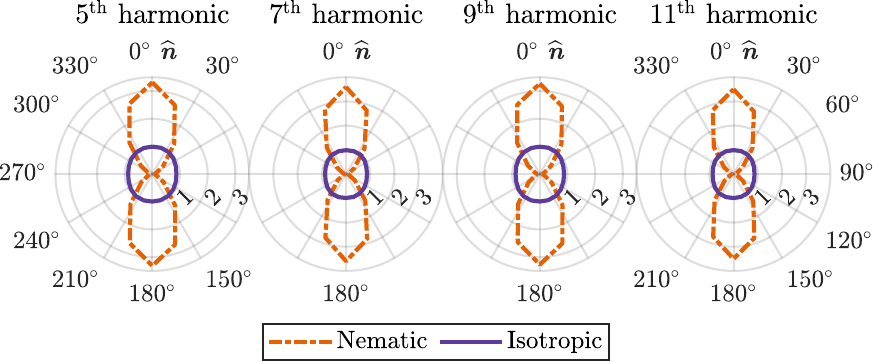}
	\caption{Emitted harmonic yield as a function of polarization direction with respect to the molecular director $\bm n$ comparing the nematic (isotropic) phase at a temperature of $30 ^\circ$ C ($50 ^\circ$ C) with order parameter $S=0.52$ ($S=0$). Intermolecular interactions are neglected, and harmonic yields are normalized to the isotropic response, imposing field-inversion symmetry and obtained utilizing laser parameters of Fig.~\ref{fig:hhg_spec}.} \label{fig:hhg_nematic_intra}
\end{figure}
Since the molecular structural characteristics of the disordered liquid phase and the partially ordered nematic phase bear many similarities to the orientational averaged and aligned gas phase results, we take our starting point in a similar setting. We initially consider a collection of non-interacting molecules with the semi-empirical tight-binding approach of Sec.~\ref{sec:2}. Doing so, grants a polarization-resolved harmonic spectroscopy of Fig.~\ref{fig:hhg_nematic_intra}. As expected, the two phases offer spectroscopic signatures, which directly reflect the statistical distribution of the independent molecular emitters. The nematic harmonic yield reaches a maximum along the nematic director, reminiscent of the aligned gas phase results of Fig.~\ref{fig:hhg_gas} or the independent molecular approach of Fig.~\ref{fig:polar_plot}, where the nonlinear efficiency is constructively enhanced along the long molecular axis. A minima is retrieved for the nematic phase when driven orthogonal to the molecular director, which compensates exactly for the maxima, such that the average harmonic yield remains constant between the two phases, as prescribed by Fig.~\ref{fig:hhg_spec}~(a). The overall polarization-dependent yield reflects the fact that all individual molecules contribute with identical single-molecular response, causing the optical anisotropies to arise due to the macroscopic distribution of such. Keeping this in mind, for soft-matter systems in which single-molecular response dominates, we expect that the polarization-dependent harmonic yield can be utilized as an ultrafast nondestructive all-optical spectroscopic technique to gain insight into the order parameters or distribution functions that govern the statistical structural description of the soft-matter sample.

Compared to the experimental results of Ref.~\cite{10.1063/5.0191184,experimental_paper}, a few insights can be uncovered. The isotropic phase exhibits a good agreement with experiments, which similarly show a constant harmonic yield at varying laser polarization. Although the total harmonic yield of Fig.~\ref{fig:hhg_spec}~(b) hinted at the fact that intermolecular interactions could play a role in the harmonic generation process, the isotropic polarization-dependency exhibits no signatures to corroborate either a single or many-molecular description. To look for signatures that favor a many-molecular description of the liquid phase, in Ref.~\cite{experimental_paper} an intensity scan of the harmonic yield will be considered. However, for investigating the nonlinear anisotropies in the current work, we will, for computational efficiency, consider a noninteracting liquid phase moving forward. 

\begin{figure} 
\includegraphics[width=\columnwidth]{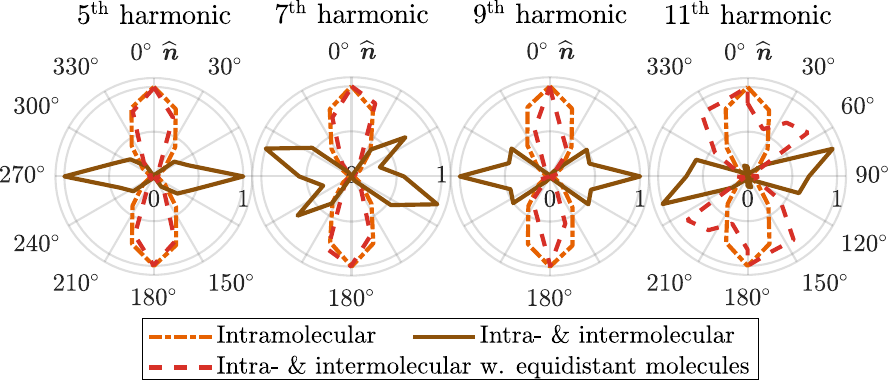}
	\caption{Normalized emitted harmonic yield as a function of polarization direction with respect to the molecular director $\bm n$ comparing a nematic phase with and without intermolecular interactions, while considering our semi-empirical structural representation or equidistant molecules. A temperature of $30 ^\circ$ C and order parameter of $S=0.52$ is considered utilizing laser parameters of Fig.~\ref{fig:hhg_spec} and field-inversion symmetry imposed on the polarization response.} \label{fig:hhg_nematic_inter}
\end{figure}

On the other hand, the experimental nematic response of Fig.~\ref{fig:1} quite remarkably exhibits a $90 ^\circ$~rotation, compared to the isolated molecular response of Fig.~\ref{fig:hhg_nematic_intra}, a feature indicating the need to go beyond an independent molecular description. Indeed, as shown in Fig.~\ref{fig:polar_plot}, the ensemble averaging of isolated molecules alone cannot lead to a polarization plot shifted by $90 ^\circ$. To go beyond the single-emiter logic, the effect of nearest-neighbor intermolecular interactions on the anisotropic spectroscopic response is investigated in Fig.~\ref{fig:hhg_nematic_inter}. Here, intermolecular interactions are observed to cause a clear rotation of the optical anisotropy, consistent across different harmonic order, to match the experimental findings. The rotation of the optical anisotropy arising from intermolecular interactions can, in its simplest case, be understood from the hybridization of molecular orbitals in Fig.~\ref{fig:homo_lumo}. On a larger scale, the macroscopic hybridization of electronic orbitals across multiple molecules can provide light-driven intermolecular electron dynamics, a precursor to the condensed matter intraband electron dynamics, which dominate harmonic emission in frequencies below the optical bandgap. In order to retrieve the rotated optical anisotropies, it is crucial to employ a description that accounts for temperature-dependent structural changes caused by both a varying order parameter and a varying density. If distributing the molecules equidistantly in both directions along and perpendicular to the nematic director, then the observed experimental anisotropy is not retrieved as observed in Fig.~\ref{fig:hhg_spec}, as harmonic emission remains maximum along the molecular director. With this in mind, we allude to the fact that the temperature-dependence of the molecular structural packing plays a key role in fortifying the relevance of intermolecular electron dynamics. For the present case of a rod-like LC, the gradual ordering of the molecules gives rise to an overall decreased center-of-mass intermolecular distance along neighboring molecules orthogonal to the nematic director, compared to those along the nematic director. As a result, the intermolecular interactions are greatly enhanced in the direction perpendicular to the nematic director as the density and order parameter increase with decreasing temperature. The strong link between the intermolecular interactions and the macroscopic molecular packing suggests that the nonlinear optical anisotropy can provide an ultrafast all-optical spectroscopic tool to track such structural properties in real-time. In essence, these conclusions are in line with the understanding of HHG in liquids of Refs.~\cite{doi:10.1073/pnas.2514825122,Mondal2026,Mondal2023}. Where Ref.~\cite{Mondal2026} see high-energy harmonics occurring due to hybridization of molecular orbitals across solvation shells, and Refs.~\cite{doi:10.1073/pnas.2514825122,Mondal2023} see diminishing electron trajectories as electrons scatter off neighboring molecules or dense solvation shells potentials. While these two contributions can coexist and compete, giving rise to different spectroscopic features, our study alludes to the hybridization of molecular orbitals being key to determining what role the intermolecular interactions take. Compared to Ref.~\cite{doi:10.1073/pnas.2514825122}, where an especially localized radial distribution function causes a scattering potential, in our case, the characteristic tendency for rod-like LC molecules to dimerize causes intermolecular hybridization that turns a corresponding peak of the localized radial distribution into a recombination site, rather than a scattering potential. We further note, that the observed features are also well in line with expectations based on molecular crystals~\cite{Bielke2025}, where a rotation of specific nonlinear anisotropies also coincide with the intermolecular axis, however for our case, the addition of structural disorder provides a more complex optical response.

By recognizing the twofold spectroscopic role of intermolecular interactions, we predict high-order harmonic spectroscopy to inherit sensitivity towards the degree of intermolecular electronic hybridization, offering an ultrafast probe into the fundamental mechanisms driving molecular bond-formation and stereochemistry.

\section{Summary and outlook}  \label{Section4}
Intending to bring the tunable LC platform into the nonlinear optical regime of attosecond physics, we have explored high-order harmonic generation across the liquid to nematic mesophase transitions in the prototypical rod-like 5CB LC. The observable central to the historical development of LC technologies, their optical anisotropy, is explored in the nonlinear regime originating from the gaseous and crystalline phases, where the underlying mechanisms are well established. A maximum nonlinear optical response is found in the gas phase if utilizing optical alignment and driving the rod-like molecules along their long molecular axis. In the crystalline phase, the nonlinear optical anisotropy is no longer linked to the molecular long-axis, but instead grants optimal nonlinear yield if targeting specific high-symmetry points of the Brillouin zone. In the most disordered of the LC phases, the liquid phase, an isotropic nonlinear optical response is retrieved in agreement with experiments. Whereas recent studies of harmonic generation in liquids and solutions have showcased a multifaceted influence of intermolecular interactions. These interactions are found in the total emitted spectrum of our model; however, the influence of such mechanisms is not critical to reproduce the polarization-dependent nonlinear isotropic response, which can be captured well with statistically oriented independent molecules. If, however, turning to the nematic LC phase, analogous to a partially aligned molecular gas, the nonlinear optical anisotropy turns to exhibit clear signatures of intermolecular dynamics. These arise in the form of a $90 ^\circ$ rotation of the direction of maximum harmonic yield, which now is maximized along the direction orthogonal to the nematic director, a signature corroborated by experiments~\cite{10.1063/5.0191184}. Through a semi-empirical description of the underlying structural properties, we find that such a change in the optical anisotropy relies on the macroscopic molecular structure responding to the change of order parameter and density that arise with temperature variations. The strong intermolecular interactions can thereby, through an hybridization of the molecular orbitals, support a significant intraband-like harmonic generation mechanisms that dominate the anisotropies of the optical response. Through the controllable structure of the LC, we thereby provide additional perspectives to the role of intermolecular interactions in harmonic generation, and showcase a sensitivity towards many-molecular electron hybridization, which could be of relevance for ultrafast imaging of fundamental chemical processes.

After unraveling how the intricate intermolecular interactions govern the anisotropicity of harmonic generation in LCs, the next step is to utilize such knowledge for spectroscopic purposes. To do so we envision to investigate how the nonlinear anisotropy changes throughout the first-order phase transition between the liquid and nematic phase. Since the optical anisotropy is governed by the molecular ordering and packing, the spectroscopic signatures should directly probe the first-order phase transition. We predict an abrupt change in the optical anisotropy across all harmonic frequencies. While the harmonic spectrum in general offers more spectroscopic information than just the anisotropies, we argue that its sensitivity to intermolecular hybridization could provide further means for resolving the change of short- and long-range order when transferring across such a critical point. In this work, we provide a LC testbed model system, whose nonlinear optical anisotropicity is predicted to be readily controlled by temperature and electric fields, solidifying the LC platform as a prominent candidate for innovative, highly-tuneable nonlinear optical elements.

\begin{acknowledgments}
The authors are grateful for valuable discussions with Caterina Vozzi, Thomas Pfeifer, and Davide Faccialà. This work was supported by the European Research Council (ERC-2024-SyG-101167294; UnMySt), the Cluster of Excellence Advanced Imaging of Matter (AIM), Grupos Consolidados y Alto Rendimiento UPV/EHU, Gobierno Vasco (IT1453-22). We acknowledge support from the Max Planck-New York City Center for Non-Equilibrium Quantum Phenomena. The Flatiron Institute is a division of the Simons Foundation. S.V.B.J. acknowledges support from the Alexander von Humboldt Foundation.
\end{acknowledgments}

\section*{Data Availability}
The first-principles electronic-structure and dynamics calculations use the open-source Octopus package~\cite{10.1063/1.5142502} (\url{https://octopus-code.org}). Input files for Octopus, as well as additional analysis scripts, may be obtained from the authors upon reasonable request.

\appendix

\section{Derivations of $A_1$, $A_2$, and $A_3$ coefficients}
\label{sec:derivation_A_coeff}

We seek to obtain the coefficients $A_1$, $A_2$, and $A_3$ in the expression below
\begin{eqnarray}
 \langle \nu_i\nu_j\nu_k\nu_l\rangle &=& A_1(\delta_{ij}\delta_{kl} + \delta_{ik}\delta_{jl} + \delta_{il}\delta_{jk}) +  \nonumber\\
 && A_2(n_in_j\delta_{kl} +\ldots) + A_3 n_i n_j n_k n_l\,,
\end{eqnarray}
where $n_in_j\delta_{kl}+\ldots$ denotes the six distinct permutations containing two director components and one Kronecker delta.
The averaging here is performed with respect to the two angles that define the molecular orientation, i.e., $\theta$ and the azimuthal angle $\phi$. For an arbitrary orientation-dependent quantity $g(\theta,\phi)$, the average is therefore defined as
\begin{equation}
\langle g\rangle =
\int_0^\pi d\theta\, f(\theta)\sin(\theta)
\frac{1}{2\pi}\int_0^{2\pi} d\phi\, g(\theta,\phi)\,.
\nonumber
\end{equation}
In order to compute the coefficients, we compute first the symmetry-axis component $\langle \nu_z^4\rangle$ with $\nu_z = \cos(\theta)$.
Expanding in Legendre polynomials $ \cos^4(\theta) = \frac{8}{35}P_4(\cos(\theta)) + \frac{4}{7}P_2(\cos(\theta)) + \frac{1}{5}$ we find
\begin{equation}
 \langle \nu_z^4\rangle = \frac{8}{35}S_4 + \frac{4}{7}S + \frac{1}{5}\,,\nonumber
\end{equation}
where we introduced the fourth-order scalar parameter
\begin{equation}
 S_4 = \langle P_4 (\cos(\theta)) \rangle. \nonumber
\end{equation} 
Considering the mixed terms, we have
\begin{equation}
 \langle \nu_z^2\nu_x^2 \rangle = \langle \cos^2(\theta)\sin^2(\theta) \cos^2(\phi)\rangle = \frac{1}{2}\langle \cos^2(\theta)\sin^2(\theta)\rangle\,, \nonumber
\end{equation}
where the last step is obtained from averaging over $\phi$.
Using the relation
\begin{equation}
 \cos^2(\theta)\sin^2(\theta) = \cos^2(\theta) - \cos^4(\theta)\,,\nonumber
\end{equation}
along with Legendre expansion $\cos^2(\theta) = \frac{2}{3}P_2(\cos(\theta)) + \frac{1}{3}$, we obtain that
\begin{align}
\langle \nu_z^2\nu_x^2 \rangle =& \frac{1}{2}\Big( \frac{1}{3} + \frac{2}{3}S \Big) - \frac{1}{2}\Big( \frac{1}{5} + \frac{4}{7}S + \frac{8}{35}S_4\Big), \nonumber \\  =& \frac{1}{15} + \frac{1}{21}S - \frac{4}{35}S_4\,. \nonumber
\end{align}
The last in-plane component required for determining the coefficients is 
\begin{equation}
 \langle \nu_x^2\nu_y^2 \rangle = \frac{1}{4}\langle \sin^4(\theta) \sin^2(2\phi)\rangle = \frac{1}{8}\langle  \sin^4(\theta) \rangle\,. \nonumber
\end{equation}
We apply the relation
\begin{equation}
  \sin^4(\theta) = 1-2\cos^2(\theta) + \cos^4(\theta),\nonumber
\end{equation}
while using the previous expressions, to obtain
\begin{equation}
 \langle \nu_x^2\nu_y^2 \rangle =  \frac{1}{15} - \frac{2}{21}S + \frac{1}{35}S_4\,. \nonumber
\end{equation}
We now relate these terms to the parameters $A_1$, $A_2$, and $A_3$, by recognizing that
\begin{eqnarray}
 \langle \nu_x^2\nu_y^2 \rangle &=& A_1\nonumber\\
 \langle \nu_z^2\nu_x^2 \rangle &=& A_1 + A_2 \nonumber\\
 \langle \nu_z^4 \rangle &=& 3A_1 + 6A_2 + A_3\nonumber.
\end{eqnarray}
From such expression, it follows that 
\begin{eqnarray}
 A_1 &=& \frac{1}{15} - \frac{2}{21}S + \frac{1}{35}S_4 = \frac{7-10S+3S_4}{105}\nonumber\\
 A_2 &=& \frac{S-S_4}{7}  \nonumber\\
 A_3 &=& S_4\nonumber.
\end{eqnarray}
From which, we obtain
\begin{eqnarray}
 \langle \nu_i\nu_j\nu_k\nu_l\rangle = \frac{7-10S+3S_4}{105}(\delta_{ij}\delta_{kl} + \delta_{ik}\delta_{jl} + \delta_{il}\delta_{jk}) \nonumber\\
 + \frac{S-S_4}{7}(n_in_j\delta_{kl} +\ldots) + S_4 n_i n_j n_k n_l\,, \nonumber
\end{eqnarray}
which gives the orientationally averaged expression for the third-order susceptibility.

\section{Expression for the third order polarization}
\label{app:P3}

It is possible to express the third order polarization as
\begin{equation}
 P^{(3)}_i = \sum_{jlk} \chi_{ijkl}^{(3)} E_jE_kE_l\,.\nonumber
\end{equation}
Using the definition of the electric field, we obtain the expression of the third-order polarization in the polarization plane, i.e. given by the xxxx, zzzz, xxzz, and zzxx components,
\begin{eqnarray}
  P_x^{(3)}(\theta)  &=&  \Big(\chi_{xxxx}+ 3\chi_{xxzz} \Big)E_0^3 \nonumber\\
  &=& \chi_{\parallel}^{(3)}E_0^3 \Big( \langle \nu_x^4 \rangle\sin^3(\theta) + 3 \langle \nu_z^2\nu_x^2 \rangle \cos^2(\theta)\sin(\theta)\Big)\,,\nonumber
\end{eqnarray}
and
\begin{eqnarray}
  P_z^{(3)}(\theta)  &=& \Big(\chi_{zzzz} + 3\chi_{zzxx} \Big)E_0^3  \nonumber\\
  &=& \chi_{\parallel}^{(3)}E_0^3 \Big(\langle \nu_z^4 \rangle\cos^3(\theta) + 3\langle \nu_z^2\nu_x^2 \rangle\cos(\theta)\sin^2(\theta) \Big)\,. \nonumber
\end{eqnarray}
Combining such expressions gives 
\begin{align}
 \vec{P}^{(3)}(\theta)& = 
\chi_{\parallel}^{(3)}E_0^3 \times \nonumber \\ &\Bigg( 3\langle \nu_z^2\nu_x^2 \rangle 
\begin{pmatrix}
\cos^2(\theta)\sin(\theta) \\
\cos(\theta)\sin^2(\theta)
\end{pmatrix}
+
\begin{pmatrix}
\langle \nu_x^4 \rangle \sin^3(\theta) \\
\langle \nu_z^4 \rangle \cos^3(\theta) 
\end{pmatrix}
\Bigg)\,.\nonumber
\end{align}
To relate such polarization dependency towards the order parameters, we use the expression of $\langle \nu_i\nu_j\nu_k\nu_l\rangle$ in terms of $A_1$, $A_2$, $A_3$, which gives, considering first the component along the director for compactness,
\begin{align}
 &3\langle \nu_z^2\nu_x^2 \rangle\cos(\theta)\sin^2(\theta) + \langle \nu_z^4 \rangle \cos^3(\theta)\nonumber\\  &= (3A_1 + 3A_2)\cos(\theta)\sin^2(\theta) + (3A_1 + 6A_2 + A_3)\cos^3(\theta), \nonumber\\
 &= 3(A_1 + A_2)\cos(\theta) + (3A_2 + A_3)\cos^3(\theta)\,. \nonumber
\end{align}
Since $\langle \nu_x^4 \rangle = 3A_1$, we similarly obtain
\begin{align}
 \langle \nu_x^4  \rangle\sin^3(\theta) + &3 \langle \nu_z^2\nu_x^2 \rangle \cos(\theta)^2\sin(\theta)\nonumber\\
 &= 3A_1 \sin^3(\theta)  + 3(A_1+A_2)\cos(\theta)^2\sin(\theta), \nonumber\\
 &= 3A_1 \sin(\theta) +3A_2\cos(\theta)^2\sin(\theta)\,, \nonumber
\end{align}
which leads to the expression of the polarization vector
\begin{align}
 \vec{P}^{(3)}(\theta) =& 
\chi_{\parallel}^{(3)}E_0^3 \times \nonumber \\
&\begin{pmatrix}
3[A_1 + A_2\cos(\theta)^2] \sin(\theta)  \\
[3(A_1 + A_2) + (3A_2 + A_3)\cos^2(\theta)]\cos(\theta) 
\end{pmatrix} .\nonumber
\end{align}
\section{Generalization to high-order harmonics}\label{sec:higher-order}

\begin{table*}[t]
	\begin{tabular}{|c|c|c|c|c|c|c|c|c|c|c|c|c|c|c|c|c|c|c|c|c|}
		\hline
		Mol & N1 & C1 & C2 & C3 & C4 & C5 & C6 & C7 & C8 & C9 & C10 & C11 & C12 & C13 & C14 & C15 & C16 & C17 & C18   \\
		\hline
		x [\AA]& -7.57 & -6.42 & -5.01 & -4.44 & -3.07 & -2.22 & -0.76 & -0.05 & 1.32 & 2.03 & 1.32 & -0.05 & 3.53 & 4.31 & 5.83 & 6.62 & 8.13 & -2.80 & -4.17 \\
		\hline
		y [\AA]& -0.50 & -0.38 & -0.23 & -0.30 & -0.14 & 0.09 & 0.28 & 1.22 & 1.4 & 0.64 & -0.30 & -0.48 & 0.79 & -0.15 & -0.01 & -0.94 & -0.79 & 0.15 & -0.01   \\
		\hline
		z [\AA]& -0.61 & -0.45 & -0.27 & 1.02 & 1.19 & 0.09 & 0.28 & -0.48 & -0.29 & 0.65 & 1.41 & 1.23 & 0.81 & -0.14 & 0.00 & -0.93 & -0.78 & -1.19 & -1.37   \\
		\hline
		Mol & H1 & H2 & H3 & H4 & H5 & H6 & H7 & H8 & H9 & H10 & H11 & H12 & H13 & H14 & H15 & H16 & H17 & H18 & H19  \\
		\hline
		x [\AA]& -5.09 & -2.65 & -0.57 & 1.85 & 1.84 & -0.58 & 3.82 & 3.82 & 4.01 & 4.01 & 6.12 & 6.12 & 6.33 & 6.33 & 8.45 & 8.45 & 8.68 & -2.16 & -4.6  \\
		\hline
		y [\AA]& -0.46 & -0.16 & 1.84 & 2.15 & -0.91 & -1.23 & 0.57 & 1.83 & 0.06 & -1.19 & -0.22 & 1.03 & -0.73 & -1.98 & -1.02 & 0.24 & -1.45 & 0.3 & 0.03 \\
		\hline
		z [\AA]& 1.88 & 2.20 & -1.20 & -0.88 & 2.15 & 1.82 & 1.85 & 0.61 & -1.18 & 0.06 & 1.04 & -0.19 & -1.98 & -0.74 & 0.25 & -1.00 & -1.46 & -2.06 & -2.37   \\
		\hline
	\end{tabular}
\caption{Atomic coordinates of the 5CB structure obtained from geometry optimization.}\label{tab:coord}
\end{table*}

The approach for the $3^\mathrm{rd}$ order term of the previous appendices can be extended to higher-order processes, considering the $N^\mathrm{th}$ order susceptibility of the isolated molecule in the form
\begin{equation}
 \chi_{i_1\ldots i_{N+1}}^{(N)}(\vec{\nu}) = \chi_{\parallel}^{(N)} \nu_{i_1}\ldots \nu_{i_{N+1}}\,,\nonumber
\end{equation}
with an expression for the averaged tensor equivalent to Eq.~\eqref{equivalent}.
The $N^\mathrm{th}$ order polarization response, along the direction $i$, can be written on the generic form
\begin{equation}
 P_i^{(N)} = \chi_{\parallel}^{(N)}  \langle \nu_i(\vec{E}\cdot\vec{\nu})^N \rangle \,.
\end{equation}
where the orientational average contains both the polar angle that defines the projection along the director and the azimuthal angle $\phi$ around it, as defined in App.~\ref{sec:derivation_A_coeff}.
For arbitrary $N$, an exact treatment would require all orientational moments up to rank $N+1$. We therefore introduce a mean-angle closure in which the polar-angle dependence is represented by its second moment, while the azimuthal average is retained explicitly.
The molecular orientation is approximated as $\nu_x \approx C\cos(\phi)$, $\nu_y  \approx C\sin(\phi)$, and $\nu_z  \approx \mu$, with $\mu^2 = \langle \nu_z \nu_z  \rangle = \frac{1 +2S}{3}$ (from Eq.~\ref{eq:1}), and $C=\sqrt{1-\mu^2}$. The parameter $\mu$ is therefore the effective projection of the molecular axis along the director, while $C$ is the corresponding transverse projection fixed by normalization. This closure reproduces the third-order result once the same $S_4(S)$ decoupling is used, but it should not be interpreted as an exact orientational average for arbitrary order.
We are now left with the azimuthal integral over $\phi$  and the angular dependence due to the polarization angle $\theta$ formed by the electric field and the director of the LC. We make use of the binomial identity
\begin{align*}
 (\vec{E}\cdot\vec{\nu})^N &= E_0^N(\sin(\theta) \nu_x + \cos(\theta)\nu_z)^N \\ &
 = E_0^N\sum_{k=0}^N 
 \begin{pmatrix}
N \\
k
\end{pmatrix}
\sin^k(\theta) \nu_x^k \cos^{N-k}(\theta) \nu_z^{N-k}\,,
\end{align*}
where we used the binomial theorem to obtain the second equality.

With $\phi$-averaging, and the above definitions, we arrive to 
\begin{eqnarray}
 P_x^{(N)}&(\theta)& = \chi_{\parallel}^{(N)}  \frac{1}{2\pi}\int_{0}^{2\pi} \nu_x  (\vec{E}\cdot\vec{\nu})^N  d\phi \nonumber\\
 &=& \chi_{\parallel} E_0^N \sum_{k=0}^N 
 \begin{pmatrix}
N \\
k
\end{pmatrix}
 \sin^k(\theta)  \cos^{N-k}(\theta)\mu^{N-k} \langle \nu_x^{k+1}\rangle_{\phi}. \nonumber
\end{eqnarray}
From $\nu_x$, we obtain azimuthal averaging over powers of cosines. For all odd $m$, we have $\langle \cos^m\phi\rangle_\phi = 0$, while for even $m=2j$, we instead have the relation $\langle \cos^{2j}\phi\rangle_\phi = \frac{(2j)!}{2^{2j}(j!)^2}$.
We can therefore rewrite the transverse component as
\begin{eqnarray}
 P_x^{(N)}(\theta) &=& \chi_{\parallel}^{(N)} E_0^N \sum_{j=1}^{\lfloor\frac{N+1}{2}\rfloor}  \left[\begin{pmatrix} 
N \\
2j-1
\end{pmatrix} \sin^{2j-1}(\theta)\times \right. \nonumber\\ 
 &&
\left.
   \cos^{N-2j+1}(\theta)\mu^{N-2j+1} C^{2j}\frac{(2j)!}{2^{2j}(j!)^2}\right]\,,
 \label{eq:PN_x}
\end{eqnarray}
where $\lfloor x\rfloor$ denotes the floor of $x$.
We obtain $P_z^{(N)}(\theta) $ in a similar way
\begin{eqnarray}
 P_z^{(N)}(\theta) &=& \chi_{\parallel}^{(N)} E_0^N \sum_{j=0}^{\lfloor\frac{N}{2}\rfloor} \left[
 \begin{pmatrix}
N \\
2j
\end{pmatrix} \sin^{2j}(\theta)  \times \right. \nonumber \\
&&
\left.
  \cos^{N-2j}(\theta)\mu^{N-2j+1} C^{2j}\frac{(2j)!}{2^{2j}(j!)^2}\right].
 \label{eq:PN_z}
\end{eqnarray}
These generic expressions provide, within the approximations described in this section, the nonlinear polarizations to order $N$. They are used in Fig.~\ref{fig:polar_plot} to compare the mean-angle closure with the explicit third-order result and with the higher-order polarization dependencies.

\section{Structural geometry optimized atomic coordinates} \label{App:P4}

In Tab.~\ref{tab:coord} the converged molecular structure of 5CB is given, as obtained from the first-principles methods of Sec.~\ref{sec:th} and illustrated in Fig.~\ref{fig:geometry_optim}.

%

\end{document}